# Riemann Zeroes from a Parametric Oscillator analyzed with Adiabatic Invariance, Hill´s Equation and the Least Action Principle

**Eduardo Stella and Celso L. Ladera**
**Departamento de Física, Universidad Simón Bolívar**
**A.P. 89000, Caracas 1080-A, Edo. Miranda, Venezuela**

***Corresponding author*: clladera@usb.ve**

**Abstract**. Adiabatic Invariance (AdI), Hill´s Equation formalism (HEF), and the Least Action Principle (LAP), three relevant tools of theoretical physics are here separately applied to a one- dimensional parametric oscillator of time-variable frequency that depends on an integer parameter $\lambda$. This oscillator is subjected to a perturbation which is a functional of the modulus of Riemann Zeta Function (RZF) times an oscillatory function, expecting that nontrivial zeroes could be obtained, in the critical strip, either (i) by optimizing the oscillator AdI, (ii) verifying the Magnus-Winkler equation associated to the oscillator Hill´s equation, (iii) from evaluating the Action integral of the perturbed oscillator. The optimum value of parameter $\lambda$ is firstly obtained when applying the AdI formalism, and we find that the three formalisms do lead to parametric oscillator states giving the RZF zeroes in arbitrary finite intervals on the critical line $(\sigma=0.5,\tau)$. The parameter $\lambda$ is later replaced by a random integer-valued function $\lambda(\sigma)$ written in terms of twin primes, actually defining an infinite set of perturbed oscillators of different Lagrangian. Nonetheless, when applying the LAP to these oscillators we still get those Riemann zeroes in the critical line at $\sigma_c=0.5$, in spite of both $(\sigma,\tau)$ varying simultaneously when evaluating the Action integral.

## 1. INTRODUCTION

Nontrivial Riemann zeroes have been receiving attention from physicists since the pioneering and ingenious work of van der Pohl [1], and notable increasing attention in the last two decades, when widely different mathematical formalisms have been used in optics, mechanics, quantum physics and even advanced quantum mechanics, to investigate relations of RZF to physics, particularly relations to obtain the zeroes on the critical line [1-11]. We here explore theoretically the potential of three mathematical-physics tools for obtaining Riemann zeroes by applying them to study a model of parametric oscillator, hoping to provide eventual novel paths for experiments that would be done to get such zeroes as in [1, 9].

The important concept of AdI [12-16] has been extensively used in theoretical physics *e.g.* in the development of the old quantum theory by Ehrenfest **[**12**],** in the study of oscillators with variable frequency by Landau and Lifshitz **[**13**]**, and in the analysis of the motion of particles in plasma physics [15, 16]. In Section 3 we apply AdI to a non-relativistic parametric oscillator of time-variable frequency which motion is perturbed, that perturbation being a functional of the modulus of RZF. In the definition of the time variable angular frequency $\Omega(t;\lambda,T)$ of this parametric oscillator two parameter $\lambda$ *and T* are introduced (Section 2). The fundamental concepts, and the basic quotient relation of AdI [13] then allows to define a *merit figur*e $Q_0$ (in Section 3) for evaluating the validity of such invariance, as the RZF-modulus varies *w.r.t.* ordinate $\tau$ of the complex plane $(\sigma,\tau)$, in arbitrarily chosen intervals on the *critical line*. A previous evaluation of this figure $Q_0$ allows to find, in an *auto-consistent* mode, the optimum, or proper, value of the integer parameter $\lambda$ (Section 3). Once this optimum value of this parameter $(\lambda=2)$ is obtained (section 3), further plotting the index $Q_0$ *vs.* $\tau$ allows to obtain the nontrivial roots of RZF located on any chosen interval of the critical line.

Introduced by G. W. Hill (1877) to study the rotation of the lunar perigee, the great importance of Hill´s equation in theoretical physics **[**17-22**]** stems from its applications to a number of apparently unrelated physical phenomenae, *e.g.* forced oscillations by solution perturbations, parametric amplification, chaos from switched-capacitor circuits [18**]**. Hill´s equation has also triggered the interest of mathematicians for physics, and a number of theorems and properties based on that equation constitute now a relevant formalism (HEF), that includes a useful theorem on a second order nonlinear differential equation found by Magnus and Winkler [21] which solution is the product of two linearly independent (LI) solutions of Hill´s equation. This nonlinear equation is applied in Section 4 to study anew the RZF perturbed parametric oscillator defined in Section 2. The key relation, between the solutions of the Hill differential equation and its wronskian, allows to define a second merit figure that again points to the Riemann zeroes in finite intervals of the critical line.

The Least Action Principle (LAP), in fact the principle of *Stationary Action,* the simplest and more useful principle of classical mechanics – its application extending to all areas of physics including general relativity and quantum mechanics, and even to chemistry [13, 23-26] – is applied (Section 5) to the perturbed parametric oscillator, and the stationary values of the Action are obtained. To the effect, the Action $J$ of the physical system is calculated using a Lagrangian which variational perturbation is the functional of the modulus of RZF. This time, it is the minimae of the Action that give the zeroes RZF along finite intervals of the critical line.

In Section 6 we develop a new procedure to obtain the Riemann zeroes, once again applying the LAP, yet this time without using the particular parameter value $(\lambda=2)$ found in Section 3, using instead a valid auto-consistent mode. To the effect we define an integer-valued function $\lambda(\sigma)$ in terms of Goldbach Conjecture and twin primes [27]. The Lagrangian of the oscillator is then rewritten replacing the parameter $\lambda$ by the integer set of integer values given by function *λ(σ)* as *σ* varies in the critical strip, and a new Action integral $J_g[\lambda(\sigma)]$ is calculated and plotted versus either *σ* or *τ*. These Action plots once again show oscillator states that correspond to Riemann zeroes in finite intervals of the critical line, or the expected infimum at the abscissa $\sigma_c=0.5$ of the critical line. Using the integer-valued function *λ(σ)* amounts to be using *simultaneously* a set of Lagrangian functions, and a set of test oscillators with different parameters *λ´s*, yet our formalism still giving us the expected and correct Riemann zeroes. This work addresses the application of those three theoretical-physics tools to show that the heights *τ* of nontrivial Riemann zeroes in finite intervals of the critical line, can be rigurously obtained from a particular physical system (an oscillator) with modest accuracy as in [8, 9]; this work *is not aimed* to develop computational, or numerical, procedures whatsoever to determine those zeroes heights with high accuracy (≥ 8-10 decimals) as printed in well-known tables [28]

## 2. DEFINITION AND MOTION EQUATION OF THE VARIABLE FREQUENCY PARAMETRIC OSCILLATOR

### 2.1 AN APPROXIMANT TO RZF

In this work we consider the RZF to be given - in the strip $\sigma\in[0,1]$ of the complex plane (σ,τ) - by the *N-th* order approximant [29]:

$$\xi(\sigma,\tau;N)=\frac{1}{\left[1-2^{1-(\sigma+i\tau)}\right]}\left[\sum_{n=1}^{N}\frac{(-1)^{n-1}}{n^{\sigma+i\tau}}\right]\ , \tag{1}$$

that for *N=1000* gives values of the nontrivial roots of that function with an accuracy of at least two decimals. For instance, the well-known first *positive* nontrivial Riemann zero lies at *τ =1413472514*, on the critical line (abscissa values *σ=0.5*); while the modulus *S* of the approximant Eq (1) at that point gives:

$$S(\sigma,\tau)=Modulus[\xi(\sigma,\tau;1000)]=|\xi(0.5,14.13472514;1000)|$$

$$-=0.006659673336737. \tag{2}$$

### 2.2 Definition of the parametric oscillator of time-variable frequency

The present study was done by separately applying AdI, HEF and the LAP, in that order, to a non-relativistic parametric oscillator of time-variable frequency. We begin by defining the following time-dependent auxiliary function denoted *ω* (plotted in Fig. 1):

$$\omega(t;\lambda,T)=\frac{-2\pi}{T}\left[1+\frac{\cos\left(\frac{2\pi t}{T}\right)\left[\sin\left(\frac{2\pi t}{T}\right)\right]^{\lambda}}{\lambda}\right]; \tag{3}$$

that depends on two parameters $T$ and $\lambda$, the latter being a dimensionless *positive* integer that shall be seen to play an important role when applying the AdI formalism to the parametric oscillator (in Section 3). $T$ is a time period which value shall be conveniently chosen as $T=\pi$, when obtaining the value of the parameter $\lambda$ that should comply with that AdI regime of the oscillator. As said above, the value of this parameter $\lambda$ shall be found in an auto-consistent mode (Section 3).

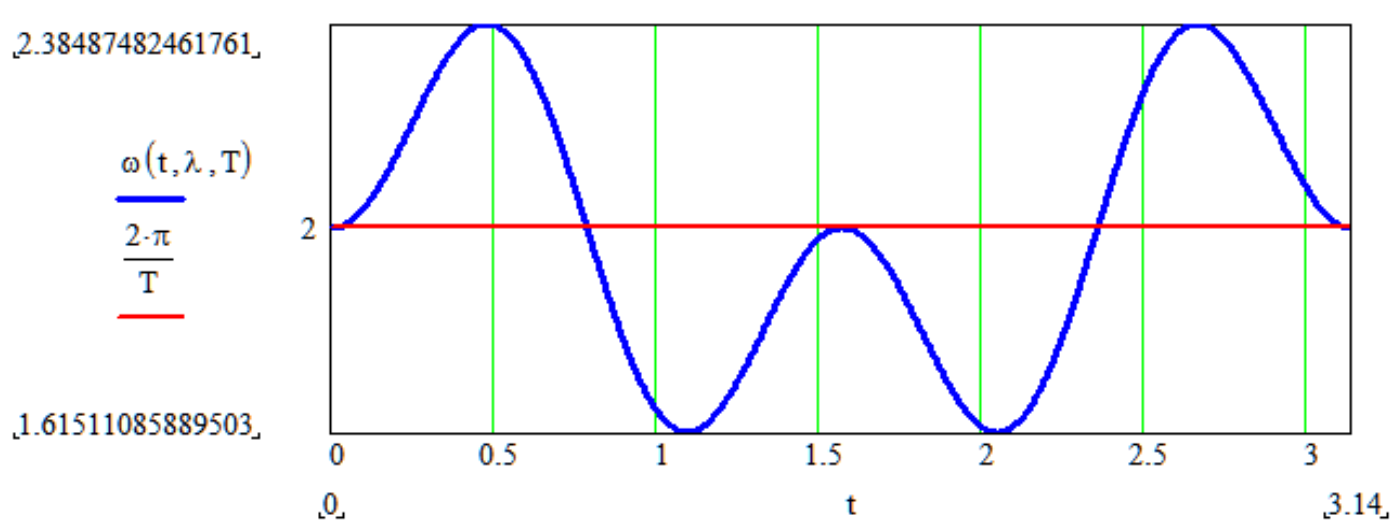

**Figure1** Graph of the auxiliary function $\omega$ in the time interval *[0,π]* and with the parameter value $\lambda=2$

In Fig. 1 we plot the function $\omega$ of Eq. (3), in the time interval *[0,π]* for the particular values $\lambda=2$ and period $T=\pi$. It should be mentioned that the definition of this function must be such to warrant that the variable frequency of the parametric oscillator, to be defined below, becomes essentially constant at $t=0$ and at $t=T$.

We also need to define a second auxiliary time-dependent function $\eta$, in terms of $\omega$, which again depends on parameters $T$ and $\lambda$,

$$\eta(t;\lambda,T)\equiv\frac{-1}{2\omega}\left(\frac{d\omega}{dt}\right)[Hz], \tag{4}$$

and plotted below in Fig. 2 for $\lambda=2$ and $T=\pi$. Note that it is a regular varying function of time. The definition of this function $\eta$ is suggested from the expected constancy of the wronskian of the Hill differential equation of the parametric oscillator (as explained in the kinematic analysis of the oscillator in Appendix 1)

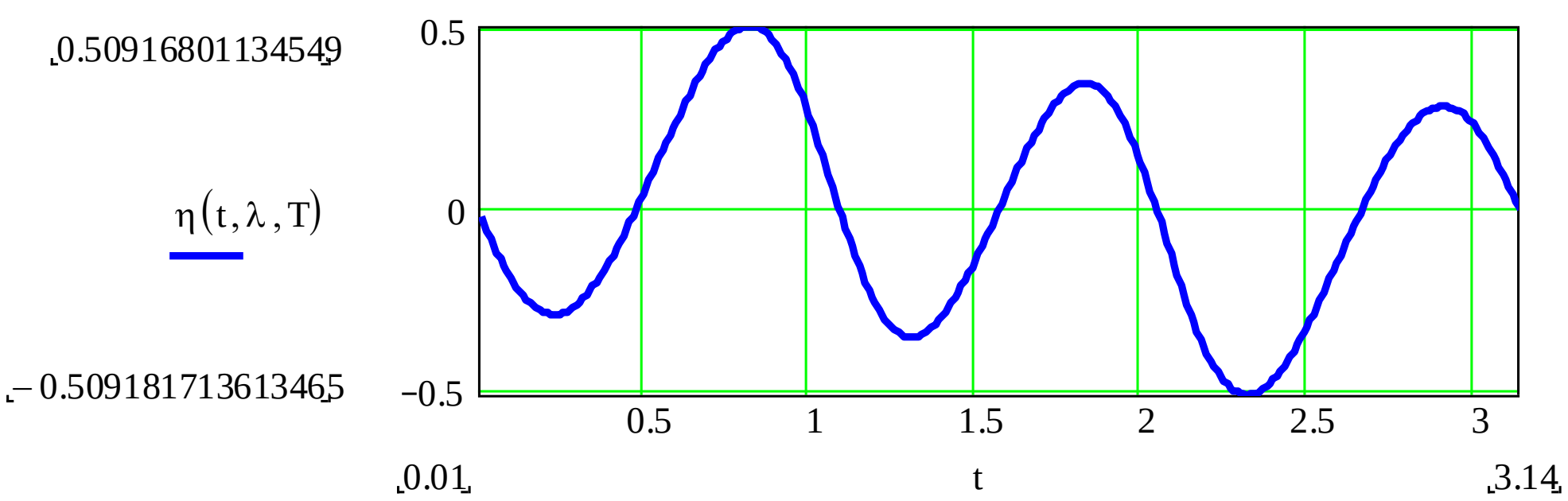

**Figure 2** Graph of $\eta\,(t;\lambda,T)$ as a function of time, with $\lambda=2$, $T=\pi$.

For the sake of simplifying the expression of the variable angular frequency $\Omega$ of the parametric oscillator we need to define a third auxiliary function $h(t;\lambda,T)$:

$$h(t;\lambda,T) = \frac{1}{\omega^2(t;\lambda,T)}\left[\frac{d\eta(t;\lambda,T)}{dt} + \eta^2(t;\lambda,T)\right]. \quad (5)$$

We may now finally define, in terms of this auxiliary function $h(t;\lambda,T)$, the time-variable angular frequency $\Omega$ of our parametric oscillator (see also Appendix 1), a function that also depends on the parameters $\lambda$ and $T$,

$$\Omega(t;\lambda,T) \equiv \omega(t;\lambda,T)\sqrt{1-h(t;\lambda,T)}. \quad (6)$$

This function is plotted in Fig. 3 below for the parameters values $\lambda = 2$ and $T=\pi$. As shown $\Omega$ is a slowly varying and almost constant angular frequency. It shall be seen that such behaviour becomes relevant when the mathematical-physics of the AdI formalism for our parametric oscillator be presented in Section 3.

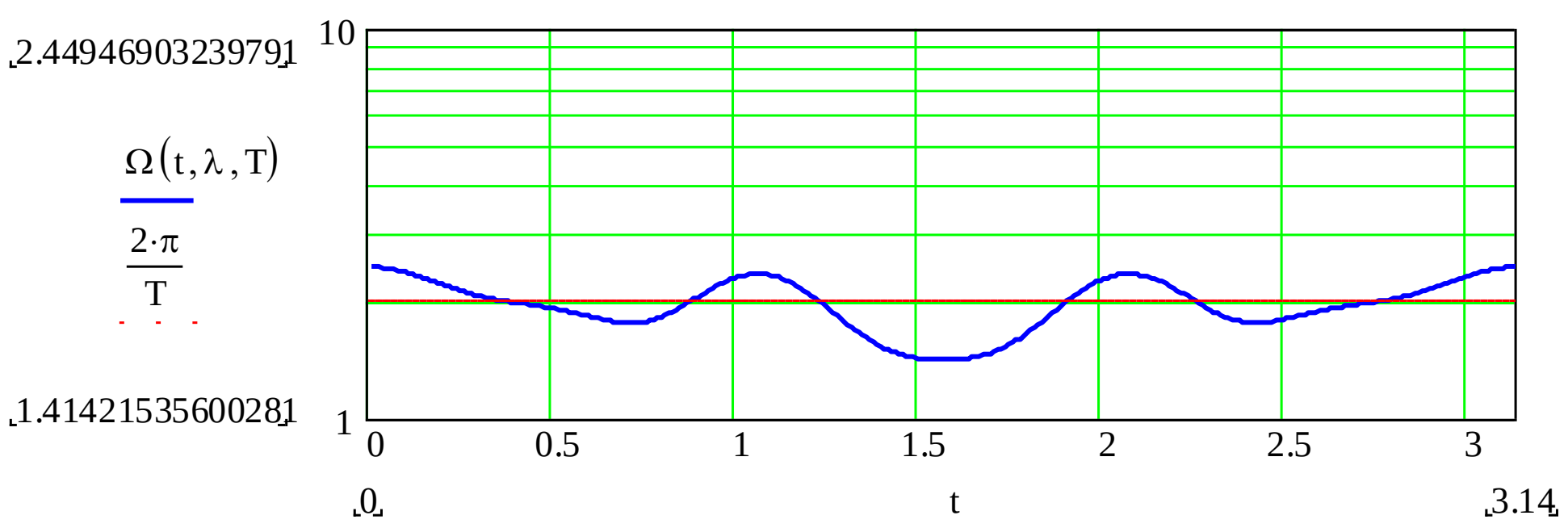

**Figure 3** Semi–log plot of the defined time dependent angular frequency $\Omega$ of the parametric oscillator for parameter values $\lambda = 2$ and $T=\pi$.

## 2.3 Hill´s equation of the parametric oscillator of time variable frequency $\Omega$

Once the time-variable frequency has been defined, we may write the Hill´s motion equation [19-21] of the parametric oscillator which mass, without loss of generality, we set as $m=1$:

$$\frac{d^2}{dt^2}X_h(t,\lambda,T) + \Omega^2 X_h(t,\lambda,T) = 0\,, \quad (7)$$

and which exact solution $X_h$, $Y_h$ we write as the oscillator functions

$$X_h(t;\lambda,T) = R(t,\lambda,T)\cos[\theta(t;\lambda,T)], \quad (9)$$

$$Y_h(t;\lambda,T) = R(t,\lambda,T)\sin[\theta(t;\lambda,T)] \quad (10)$$

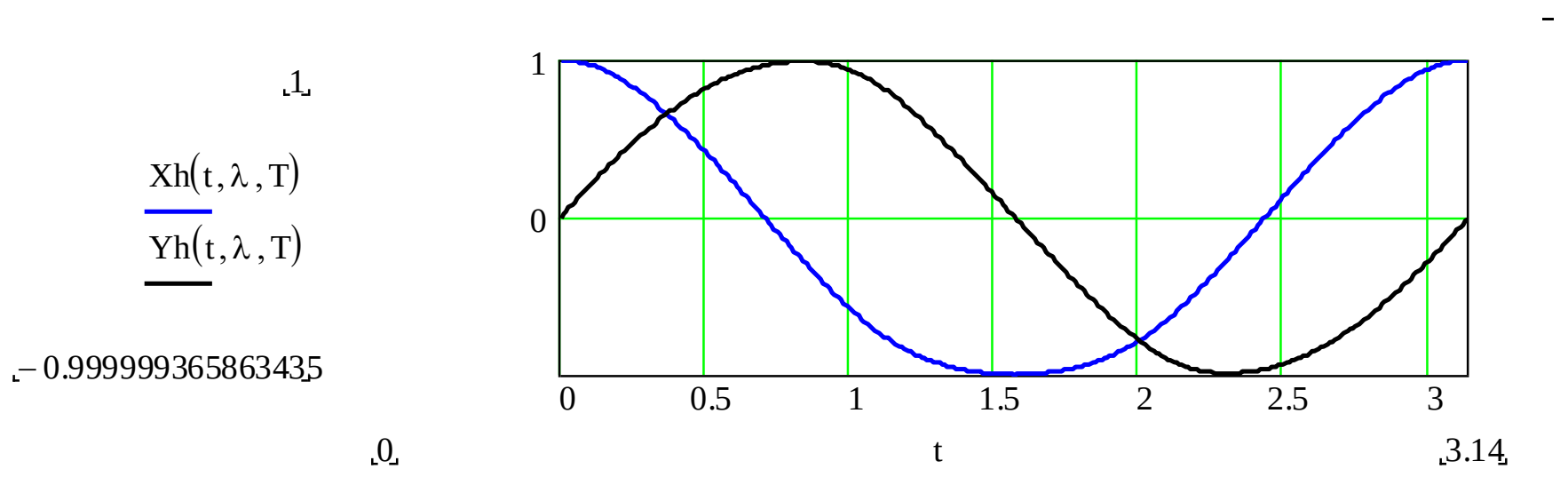

**Figure 4** Plots of the coordinates *Xh* and $Y_h$ of the solutions of Hill´s equation for the unperturbed parametric oscillator

## 3. RIEMANN ZEROES FROM THE AdI FORMALISM APPLIED TO THE PARAMETRIC OSCILLATOR OF VARIABLE FREQUENCY

We first apply the AdI formalism of classical mechanics to the following Hill-Riemann test function $X_{Hp}$ for our perturbed parametric oscillator:

$$X_{Hp}(\, t, \sigma, \tau\,, \lambda, T) \equiv X_H(t, \lambda, T)\left\{1 + S(\sigma, \tau)\sin\left[\pi\left(\frac{t}{T}\right)\right]\right\}, \tag{11}$$

which, as written, is the exact solution $X_H$ of the parametric oscillator plus a perturbation written as a functional of the modulus $S(\sigma,\tau\,)$ of RZF. As shall be seen below the AdI formalism implemented in this section aims to determine the conditions under which that perturbation be minimized. The kinetic energy $E_k$ and the potential energy $E_p$ of the perturbed parametric oscillator are respectively,

$$E_k = \frac{1}{2} m \left(\frac{d}{dt} X_{HA}(\, t, \sigma, \tau\,, \lambda, T)\right)^2 ; \tag{12}$$

$$E_p = \frac{1}{2} K(t; \lambda, P) X_{HA}(\, t, \sigma, \tau\,, \lambda, T)^2, \tag{13}$$

where the oscillator “elasticity” function $K$ van be written as follows

$$K(t; \lambda, T) = m\, \Omega(t; \lambda, T)^2 \;, \tag{14}$$

where $m=1$, as already said above.

Since we expect to apply the AdI formalism to the parametric oscillator, its variable angular frequency $\Omega$ is expected to be slowly varying. Fortunately, such is the case for our oscillator (see Fig. 4 and Eq. 4 above). The total energy $E$ of the perturbed oscillator is of course

$$E(t, \sigma, \tau\,, \lambda, T) = E_K(t, \sigma, \tau\,, \lambda, T) + E_p(t, \sigma, \tau\,, \lambda, T), \tag{15}$$

an energy that is variable in time since the perturbed parametric oscillator considered is not a conservative system. It is known [13, Section 49] that the fundamental relation of AdI may be mathematically written as:

$$\frac{1}{2\pi}\oint pdq = \frac{1}{2\pi}2\int_0^T E_k(t)dt = \frac{E(t;\sigma,\tau)}{\Omega(t,\sigma)}, \tag{16}$$

where $E$ is the total energy of the parametric oscillator, $p$ its momentum, and $\Omega$ its variable oscillation frequency. As the parameters of the oscillator vary slowly AdI must be such to warrant that the energy remains proportional to the angular frequency, *i.e.* their quotient in Eq. (16) must remain nearly constant in a period of time equal to $\pi$ [13]. To simplify writing the ensuing steps denote the quotient between the total energy energy $E$ and the angular frequency $\Omega$ in Eq. (16) – the so-called *adiabatic invariant* – as $I_{Ad}(t,\sigma,\tau,\lambda,T)$. We may then write:

$$I_{Ad}(t,\sigma,\tau,\lambda,T) = \frac{E(t,\sigma,\tau,\lambda,T)}{\Omega(t,\lambda,T)}, \tag{17}$$

Moreover denote the integral that appears at its left in Eq. (16) as:

$$I(\lambda,T) = \frac{1}{2\pi}\left\{\int_0^T 2\left[\frac{1}{2}m(X_H(t,\lambda,T))^2\right]\right\}dt \ . \tag{18}$$

Thenceforth, we may rewrite Eq. (17) in the simplified notation:

$$I(\lambda,T) \cong I_{Ad}(t,\sigma,\tau,\lambda,), \tag{19}$$

Recall that there would be particular values of the integer parameter $\lambda$ for which the angular frequency $\Omega(t,\lambda,T)$ be approximately constant –as revealed in Fig. 3– therefore warranting a valid application of AdI. Thus, in Fig. 5 we plot the variable $I_{Ad}$ (black curve), showing in effect tat it is rather close to a constant in the period $T=\pi$.

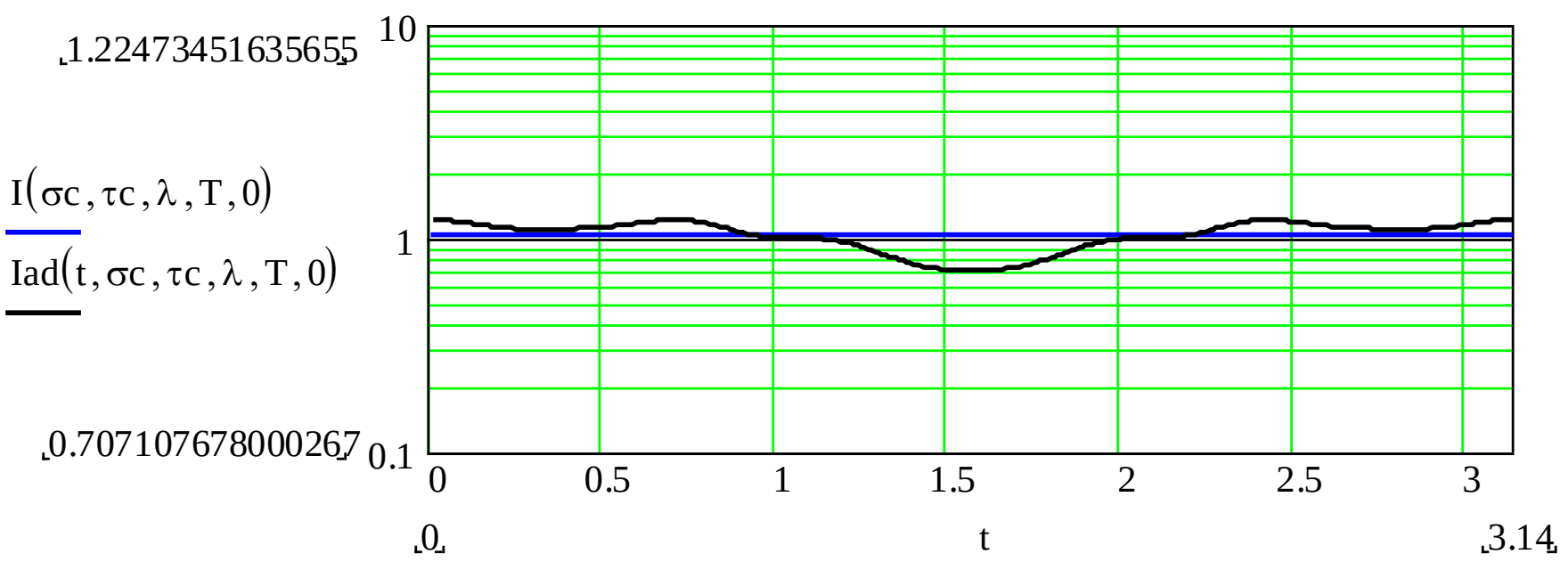

**Figure 5** Graphs of the quantities $I$ (*blue line*) and $I_{Ad}$ (*black curve*) *vs. time t* along a period $T=\pi$, for $\sigma_c=0.5$, $\tau_c=14.134$, showing $I_{Ad}$ being almost constant as expected.

We may then consider the average value $I_{Ad,ave}$ of the adiabatic invariant $I_{Ad}$ in the time interval $0 \le t \le T$:

$$I_{Ad,ave}(\sigma,\tau\,,\lambda,T\,) \equiv \frac{1}{T}\int_0^T I_{Ad}(t;\sigma,\tau\,,\lambda,T\,)dt \qquad (20)$$

After evaluating the two integrals, in Eqs (18) and (20), along a *period* $T=\pi$ we may simply write:

$$I(\lambda,T) \;=\; I_{Ad,av}(\sigma,\tau,\lambda,T) \qquad (21)$$

We shall now proceed to estimate the optimum value of the real parameter $\lambda$ introduced in Section 2 (when the auxiliary function $\omega$ was defined) in an auto-consistent way, just to warrant the correct application of the formalism of AdI. To the effect we heuristically define a merit figure, or *quality index* $Q_0(\sigma,\tau,\lambda,T)$ for any AdI application, as the following *relative error* relation of adiabatic invariance:

$$Q_0(\,\lambda,T)\equiv\left|\frac{I(\lambda,T)-I_{Ad,ave}(\sigma,\tau\,,\lambda,T)}{I(\lambda,T)}\right| \;. \qquad (22)$$

Of course this index $Q_0$ measures the relative departure between the two function values that define the AdI quotient in Eq (17), *both being dependent upon parameter* $\lambda$. As shown immediately below, this procedure allows us to find auto-consistently the proper value of the parameter $\lambda$ for AdI to hold.

Thus to get that proper value of integer parameter $\lambda$ we simply resort to evaluate, and plot index $Q_0$ for *any* particular pair chosen coordinates value $(\sigma_0,\tau_0)$, while allowing $\lambda$ to vary. That is, we assume, without loss of rigour, that the definition of index $Q_0$ in Eq. (22) holds for say two known nontrivial roots of RZF [28], *exempli gratia* $\sigma_0=0.5,\tau_0=14.134725142)$, and $\sigma_0=0.5,\tau_1=32.935061588$, while letting the integer $\lambda$ to vary from 1 to 9. The pertinent plot of index $Q_0$ is shown in Fig. 6, below:

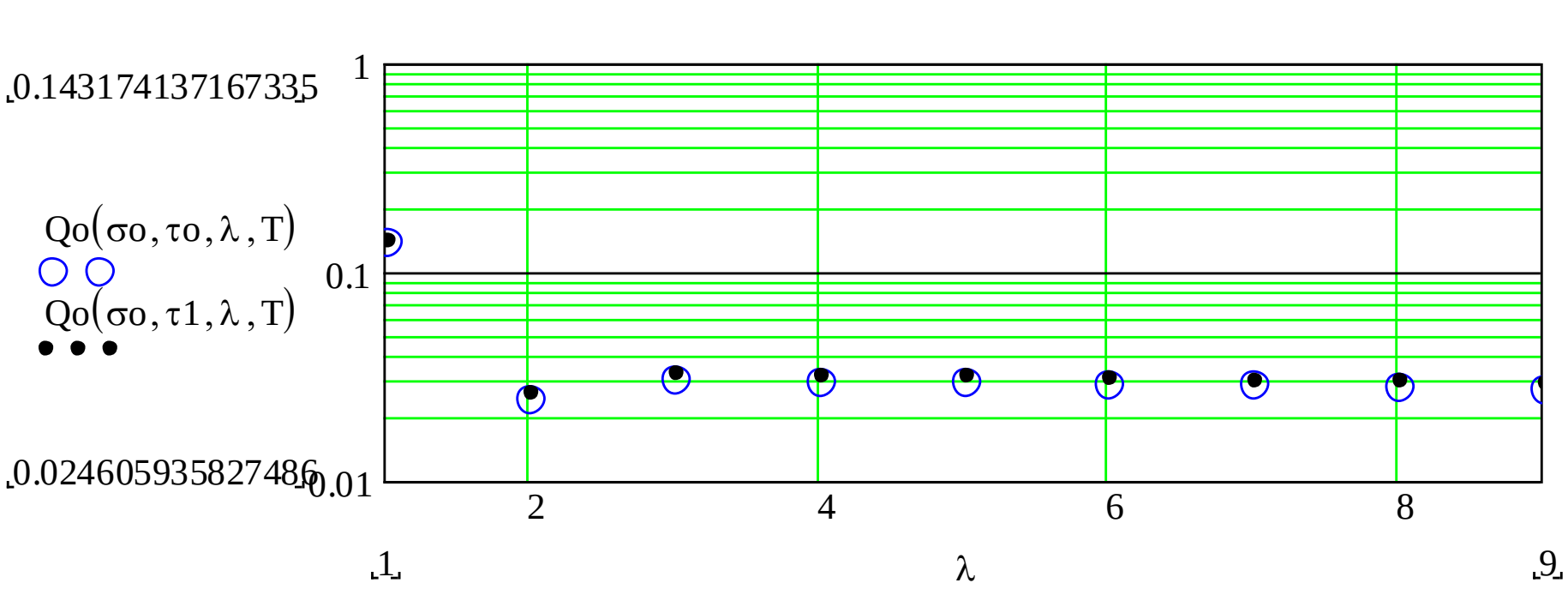

**Figure 6** Graph of the AdI index $Q_0$ *vs* parameter $\lambda$ in the interval [0, 9], for $\sigma_0=0.5$ and two different Riemann ordinates $\tau_0= 14.134725142$ (dots), $\tau_1= 32.935061588$ (circles) on the critical line: Note that the two plots superpose exactly, and have the same minimum for the parameter value $\lambda=2$.

The remarkable plot of the index $Q_0$ in Fig. 6, as a function of the parameter $\lambda$, evaluated for two pairs $(\sigma_0, \tau_0)$, $(\sigma_0, \tau_1)$ shows that in both cases it approaches zero –in fact less than $3\times10^{-2}$ – for *the same integer value* $\lambda=2$. We thus find that the AdI condition is optimum for $\lambda_c=2$ (i.e. the relative error, between the energy $E$ and the frequency $\Omega$ (Eq. (17) approaches zero), a value of the parameter $\lambda$ for which the angular frequency of the oscillator, as plotted in Fig. 3, is nearly constant.

The validity of this procedure to obtain the optimum value of our parameter $\lambda_c$ can be easily checked by re-evaluating the quality index $Q_0$ for whichever other known nontrivial roots [29] of RZF. You will find that notwithstanding the root chosen along the critical line you always get the same proper value $\lambda_c=2$, as in Fig (6). An elegant and easier way to appraise the auto-consistency of the latter method to find the $\lambda_c=2$ value, is to plot the quality index $Q_0$ versus coordinate $\tau$ of the critical line, along say the finite interval $\tau\in[90,102]$ as in Fig.7 below, or whatever any other interval you choose, for instance for the alternative finite interval $\tau\in[12,35]$, as shown in Fig. 8

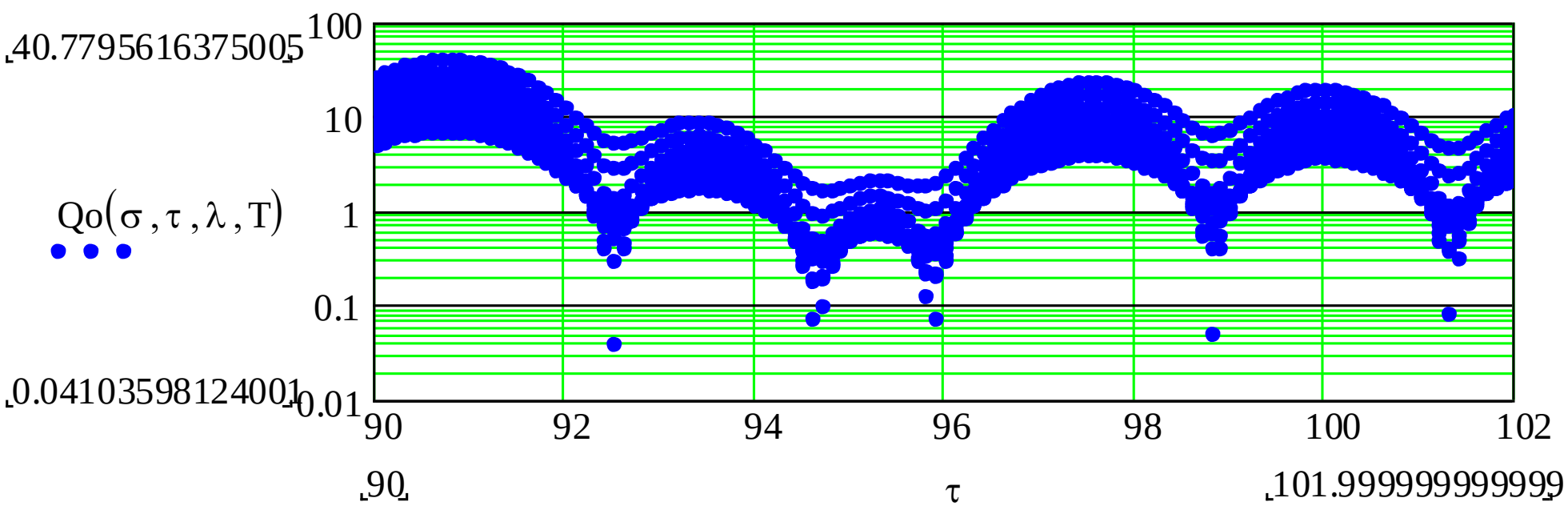

**Figure 7** Graph of the index $Q_0$ with $\tau$ in the interval *[90,102]* of the critical line, both coordinates $(\sigma,\tau)$ varying simultaneously for $\lambda=2$. The plot shows approximate values of the non-trivial Riemann zeroes on that particular interval**.**

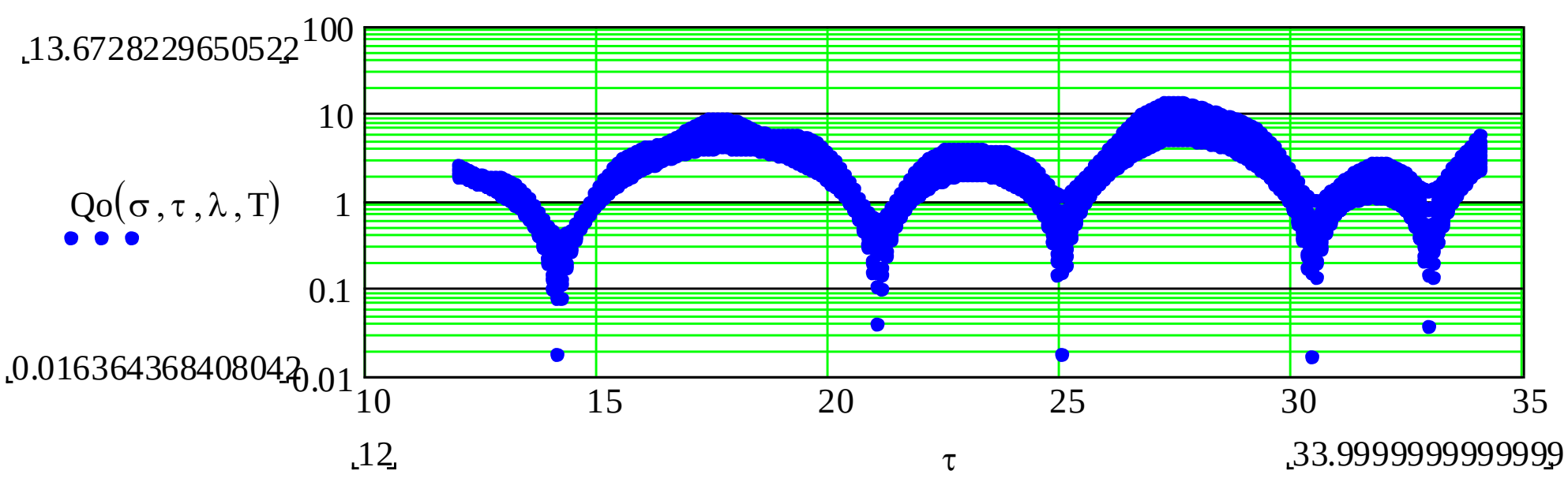

**Figure 8** Graph of the quality index $Q_0$ vs. ordinate $\tau$ in the interval *[12,35]* of the critical line, when both coordinates $(\sigma,\tau)$ vary simultaneously, and for $\boldsymbol{\lambda_c=2}$. Minimae of the index $Q_0$ show five expected Riemann zeroes, at the critical line ordinates: $\tau$ *=14.133, 21.089, 25.020, 30.413,* and *32.933* (with a maximum relative error 0.039 % *w.r.t.* currently accepted Odlyzko values [28])

For comparison, and we now list, the well-known, currently accepted, first five Riemann zeroes [28] of ordinate $\tau$ : *14.134725142*, *21.022039639*, *25.010857580*, *30.424876126*, and *32.935061588*, that belong in the same interval of the critical line used in Fig. 8 to plot the quality index $Q_0$ of adiabatic invariance *vs* $\tau$.

The last two plots (Figs.7 and 8) show that the minimae of the quality index $Q_0$ occur precisely at the nontrivial zeroes of the RZF, in spite of these two finite intervals on the critical line being widely different. We also find that plotting this index versus the abscissae $\sigma$ of the complex plane (Fig. 9)– evaluated with both coordinates $(\sigma, \tau)$ varying, and for the already found parameter value $\lambda_c=2$, the quality index $Q_0$ gets its minimum value precisely at the abscissa value $\sigma_c=0.5$ of the critical line in the strip, in coincidence with Riemann Conjecture.

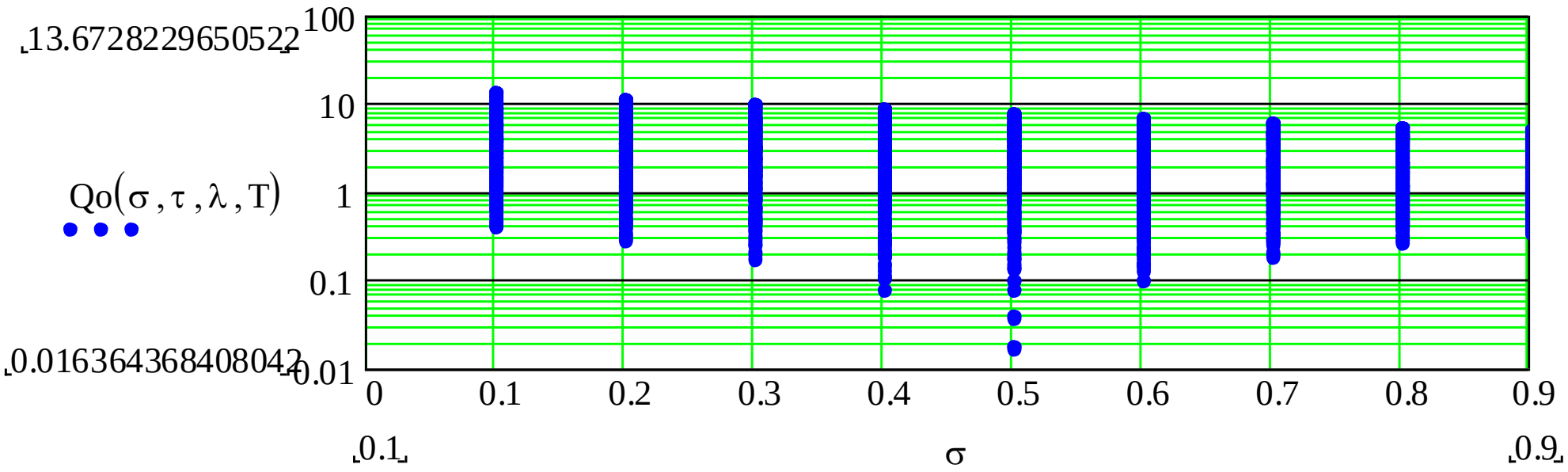

**Figure 9 Q**uality index $Q_0$ of adiabatic invariance plotted against the abscissa $\sigma$ when both coordinates $(\sigma, \tau)$ of the complex plane are allowed to vary, and with $\lambda=2$. Notice the infimum at the abscissa $\sigma_c=0.5$ of the critical line

In Section 6 we shall replace the integer parameter $\lambda$ by an integer-valued function $\lambda(\sigma)$ and then apply the LAP to the perturbed oscillator. When evaluating the Action of the oscillator using this function –that is simultaneous multiple $\lambda$ values given by this integer function $\lambda(\sigma)$– we shall still get the expected Riemann zeroes. This could also be done with the HEF, or the AdI formalism, but of course it is easier to implement it with the LAP as we shall do in Section 6.

## 4. ZEROES OF RZF FROM APPLYING HEF (MAGNUS-WINKLER THEOREM) TO THE PARAMETRIC OSCILLATOR

Consider now the Hill´s motion equation for the parametric oscillator of variable frequency introduced in Section 2:

$$\frac{d^2}{dt^2}H(t;\sigma,\tau,\lambda,\mathrm{T}) + \Omega_0^2(t;\lambda,\mathrm{T})H(t;\sigma,\tau,\lambda,\mathrm{T}) = 0\,, \qquad (23)$$

where $H\ (t;\sigma,\tau,\lambda,T)$ represents the general solution of the equation. Consider now that $X_h(t;\sigma,\tau,\lambda,T)$ and $Y_h(t;\sigma,\tau,\lambda,T)$ are two linearly independent (LI) solutions of Hill´s equation, and write its product as $Z_h$:

$$Z_h(t;\sigma,\tau,T) = X_h(t;\sigma,\tau,T\,)Y_h(\,t;\sigma,\tau,T) \tag{24}$$

An important theorem for the product $Z$ of any two *LI* solutions of Hill´s equation has been demonstrated by Magnus and Winkler **[21],** a theorem that reads**:**

"Any product such as $Z_h$ in Eq. (24) satisfies the following nonlinear second order equation:

$$\boldsymbol{Z\,\frac{d^2}{dt^2}Z - \frac{1}{2}\left(\frac{d}{dt}Z\right)^2 + 2QZ^2 = C,} \tag{25}$$

where the coefficient $Q$ must be a periodic function *i.e.* $Q(x)=Q(x+\pi)$, while $C$ is a proportional to the squared wronskian $W^2$ of the two *LI* solutions of Hill´s Equation".

That is according to Magnus and Winkler [21]:

$$C = -\frac{1}{2}[W(t;\lambda,T)]^2 \tag{26}$$

In our case, since our function $\Omega$ introduced in Section 2 has period $\pi$, we may take $Q = \Omega_0^2(t)$, and the *wronskian* for those two *LI* solutions in Eq. (24) is, by definition:

$$W(t;\lambda,T) = X_h(t;\ \lambda,T)\frac{d}{dt}Y_h(t;\lambda,T) - Y_h(t;\lambda,T)\frac{d}{dt}X_h(t;\lambda,T) \tag{27}$$

that we now plot as a function of time for $\lambda=2$ and $T=\pi$. It may be seen that the wronskian in Eq. (27) remains indeed a constant along period $T=\pi$.

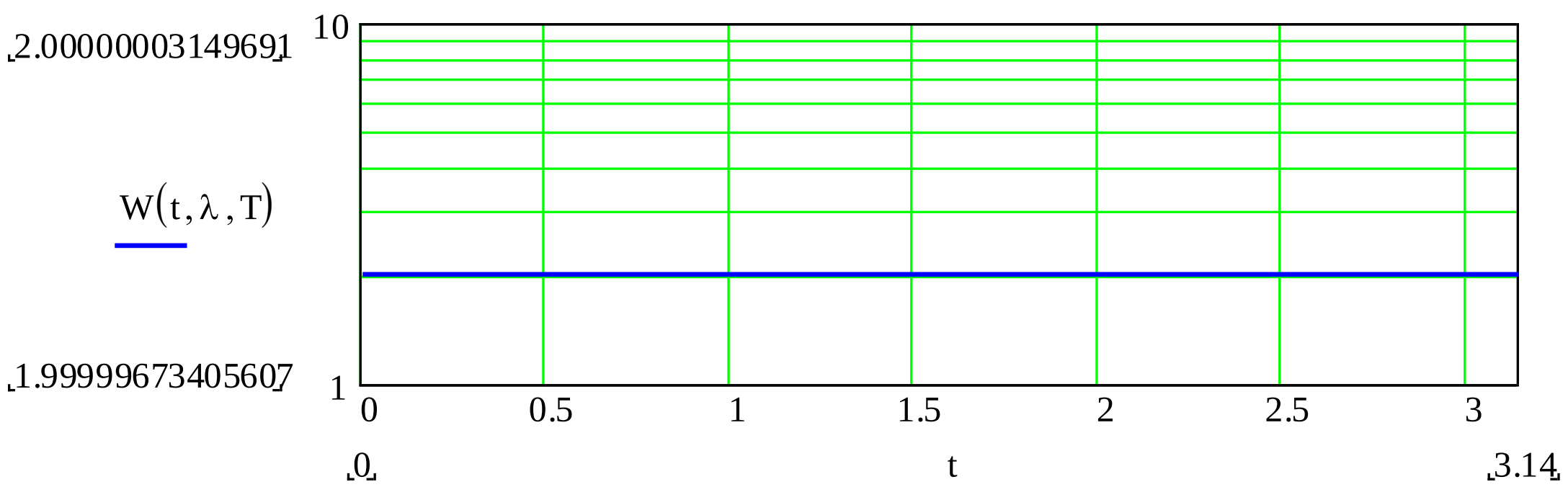

**Figure 10** Plot of the Wronskian of the two *LI* solutions, $X_h$ and $Y_h$ of Hill´s equation for $\lambda=2$ *and* $T=\pi$, showing it is in effect a constant function in a period $T=\pi$

We may now consider writing the two non-perturbed solutions of Hill´s equation for our parametric oscillator in the well known form:

$$X_h(t;\ \lambda,T) = R(t;\ \lambda,T)\cos[\theta(t;\ \lambda,T)]; \qquad (28)$$

$$Y_h(t;\ \lambda,T) = R(t;\ \lambda,T)\sin[\theta(t;\ \lambda,T)], \qquad (29)$$

while the two perturbed solutions of the parametric oscillator we may write as:

$$X_{hp}(t;\sigma,\tau,\lambda,T) = X_h(t;\lambda,T)\left[1 + S(\sigma,\tau)\sin\left[\pi\left(\frac{t}{T}\right)\right]\right], \qquad (30)$$

$$Y_{hp}(t;\sigma,\tau,\lambda,T) = Y_h(t;\lambda.T)\left[1 + S(\sigma,\tau)\sin\left[\pi\left(\frac{t}{T}\right)\right]\right], \qquad (31)$$

and consider the product of these two solutions of Hill´s Equation for the perturbed case:

$$Z_p(t;\sigma,\tau,\lambda,T) = X_{hp}(t;\sigma,\tau,\lambda,T)Y_{hp}(t;\sigma,\tau,\lambda,T) \qquad (32)$$

Applying Magnus and Winkler (M-W) theorem [**21]** to this product of solutions (in the perturbed case) we may write the quantity $C_p$, *now of course not necessarily a constant*, as proportional to the squared wronskian for the perturbed solutions:

$$C_p(t;\ \sigma,\tau,\lambda,T){=}{-}1/2\left[X_{hp}(t;\sigma,\tau\lambda,T)\frac{dY_{hp}(t;\sigma,\tau\lambda,T)}{dt} - \ Y_{hp}\ (t;\sigma,\tau\lambda,T,\rho)\frac{dX_{hp}(t;\sigma,\tau\lambda,T)}{dt}\right]^2 \qquad (33)$$

We may now rewrite the M-W theorem, Eq. (25), for the product of the perturbed solutions as:

$$EqZ(t;\sigma,\tau,\lambda,T) = C_P(t;\ \sigma,\tau,\lambda,T), \qquad (34)$$

where, the parametric function $E_qZ$, since $Q = \Omega_0^2(t)$, is given by

$$EqZ(t;\sigma,\tau,\lambda,T) = [Z_{hp}(t;\sigma,\tau,\lambda,T)\frac{d^2}{dt^2}Z_{hp}(t;\sigma,\tau,\lambda,T)\ldots$$

$$-\frac{1}{2}\left(\frac{d}{dt}Z_{hp}(t;\sigma,\tau,\lambda,T)\right)^2 + 2\Omega_0^2(t;\lambda,T)Z_{hp}(t;\sigma,\lambda,T)^2] \qquad (35)$$

In Fig.11 we have plotted functions *EqZ* (blue curve ) and $C_p$ (red curve) of Eq. (34), both in the time interval *[0, π]*, and for the particular coordinates values $\sigma_0=0.5$ $\tau_0=14.134$ of the first non-trivial zero of RZF. It may be seen that the two symmetrical graphs are, as expected, rather close to each other with a common stationary point at $\pi/2$. It may be seen that as expected function $EqZ$ follows closely function $C_p$ in that interval (showing a small maximum departure of order 0.01)

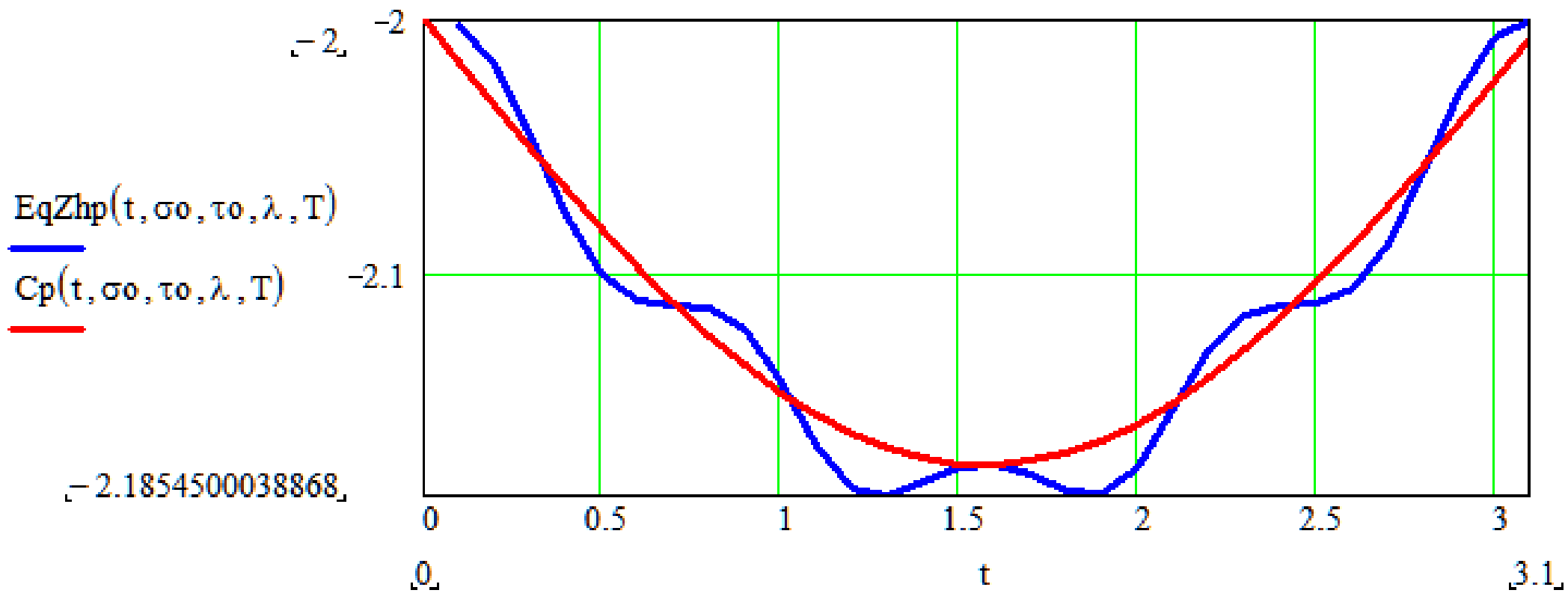

**Figure 11** Graphs of the parametric functions $EqZ$ (blue curve ) and $C_p$ (red curve), in the time interval [0, π], with $\lambda=2$, and for $\sigma_0= 0.5$ *and* the first Riemann zero at $\tau_0 \approx 14.134$

We now ought to evaluate a difference function $\Delta Z_{hp}$, defined as the departure, or difference, of the function $EqZ$ given in Eq (35) with respect to the function $C_{p.}$ given in Eq. (33), that is:

$$\Delta Z_{hp}(t;\ \sigma,\tau,\lambda,T) \equiv EqZ_{hp}(t;\ \sigma,\tau,\lambda,T) - C_p(t;\ \sigma,\tau,\lambda,T). \tag{36}$$

And let us now consider and evaluate their *relative error* $\Delta Z_{hp,rel}$ over the time interval *[0,T]*, which is a significant figure to evaluate whatever departure:

$$\Delta Z_{hp,rel}(t;\ \sigma,\tau,\lambda,T) = \left| \frac{EqZ_{hp}(t;\sigma,\tau,\lambda,T) - C_p(t;\sigma,\tau,\lambda,T)}{C_p(t;\sigma,\tau,\lambda,T)} \right|, \tag{37}$$

which we plot *vs. t* along the interval time *[0,T=π]*, in Fig. 14 below. It may be seen that the relative error $\Delta Z_{hp,rel}$ vary within a truly small range (0, ~0.01), meaning that our application of Magnus and Winkler formalism to the perturbed case is indeed valid and sound.

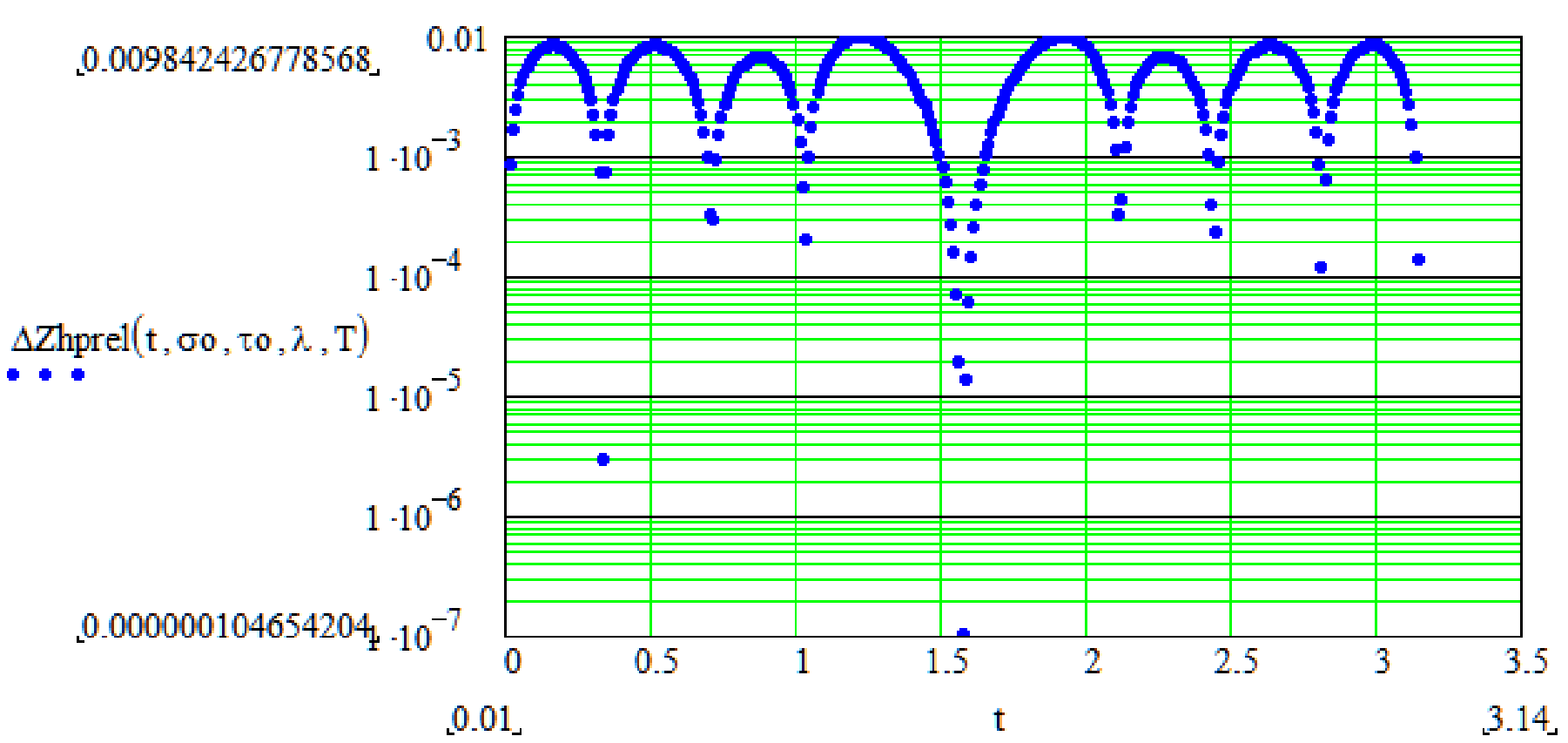

**Figure 12** Relative error of the approximation $\Delta Z_{hp,rel}$ when applying HEF (Magnus-Winkler equation) to the perturbed oscillator (maximum at 1 %)

Notwithstanding that the relative error of the aproximation $\Delta Z_{hp,rel}$ is a good figure to appraise our M-W results, a more significant figure of the quality of the application of the M-W formalism to the parametric oscillator perturbed functions, is to evaluate its *minimum-squared* difference, here denoted denote $\Delta Z_{hp,ms}$ , instead of that relative error. To the effect we apply the well-known *minimum-squared* optimization method to the squared-sum of the differences $\Delta Z_{hp}$ (for the discretization $t=nT/16$, $n$ =1, 2, 3…16), that is to evaluate:

$$\Delta Z_{hp,ms}(\sigma,\tau) = \sum_{n=1}^{16}\left[\Delta Z_{hp}\left(\frac{nT}{16};\sigma,\tau,\lambda,T\right)\right]^2 \quad (38)$$

In Fig. 14 below we have plotted such *minimum-squared difference* $\Delta Z_{hp,ms}$ as a function of the coordinate $\sigma$ along the strip (0,1):

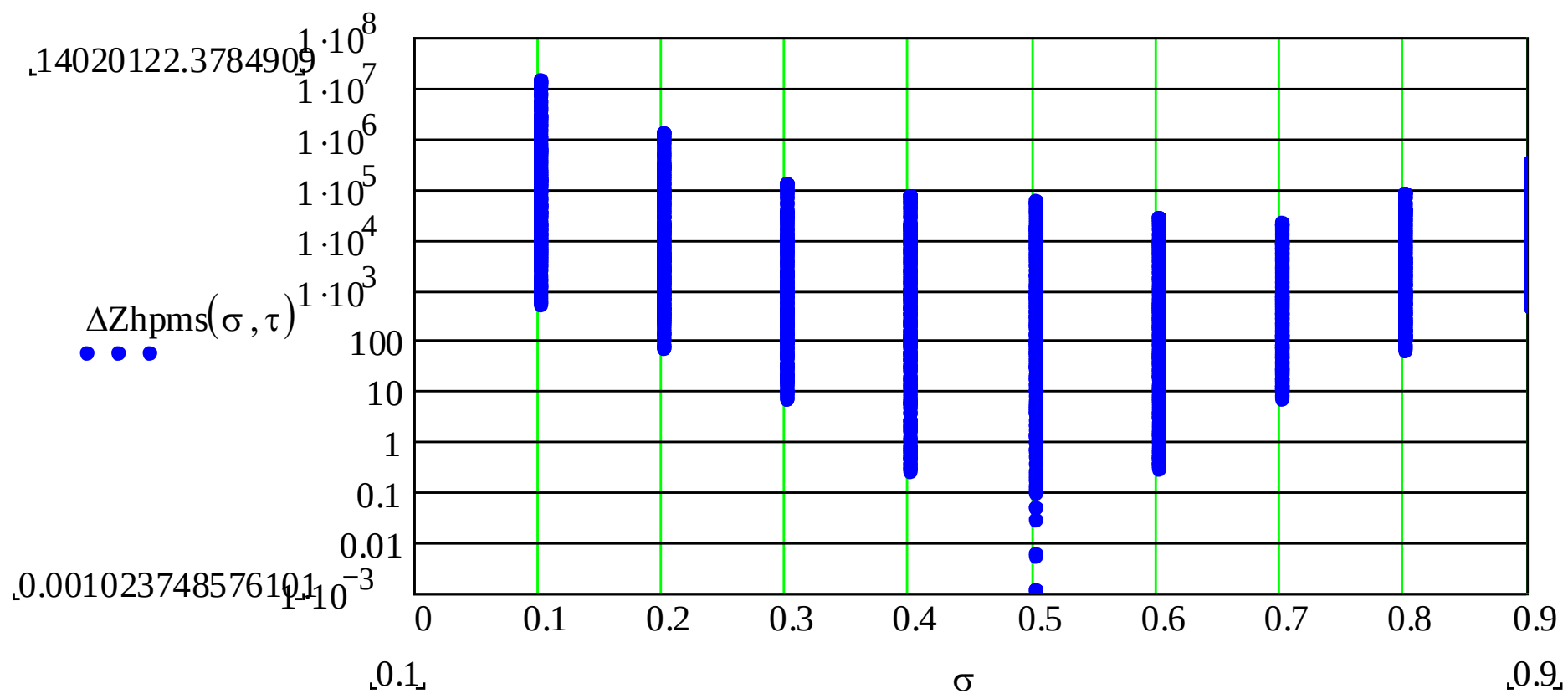

**Figure 13** Graph of the minimum-squared difference $\Delta Z_{hp,ms}$ as a function of abscissa $\sigma$ in the critical strip (0, 1) with both complex plane coordinates *(σ,τ)* varying simultaneously, giving an infimum at $\sigma=0.5$

Notably, this *minimum-squared* plot in Fig. 13 shows the difference $\Delta Z_{hp,ms}$ to have its infimum (~$10^{-3}$) at the coordinate $\sigma=0.5$, *i.e.* at the abscissa of the RZF critical line.

Moreover, we may also plot this *minimum-squared* difference $\Delta Z_{hp,ms}$ versus the ordinate $\tau$ of the critical line, in the interval $\tau\in[10, 35]$, again letting $\sigma$ and $\tau$ to vary simultaneously (note the manifold of curves in Fig. 14, due to the varying $\tau$),

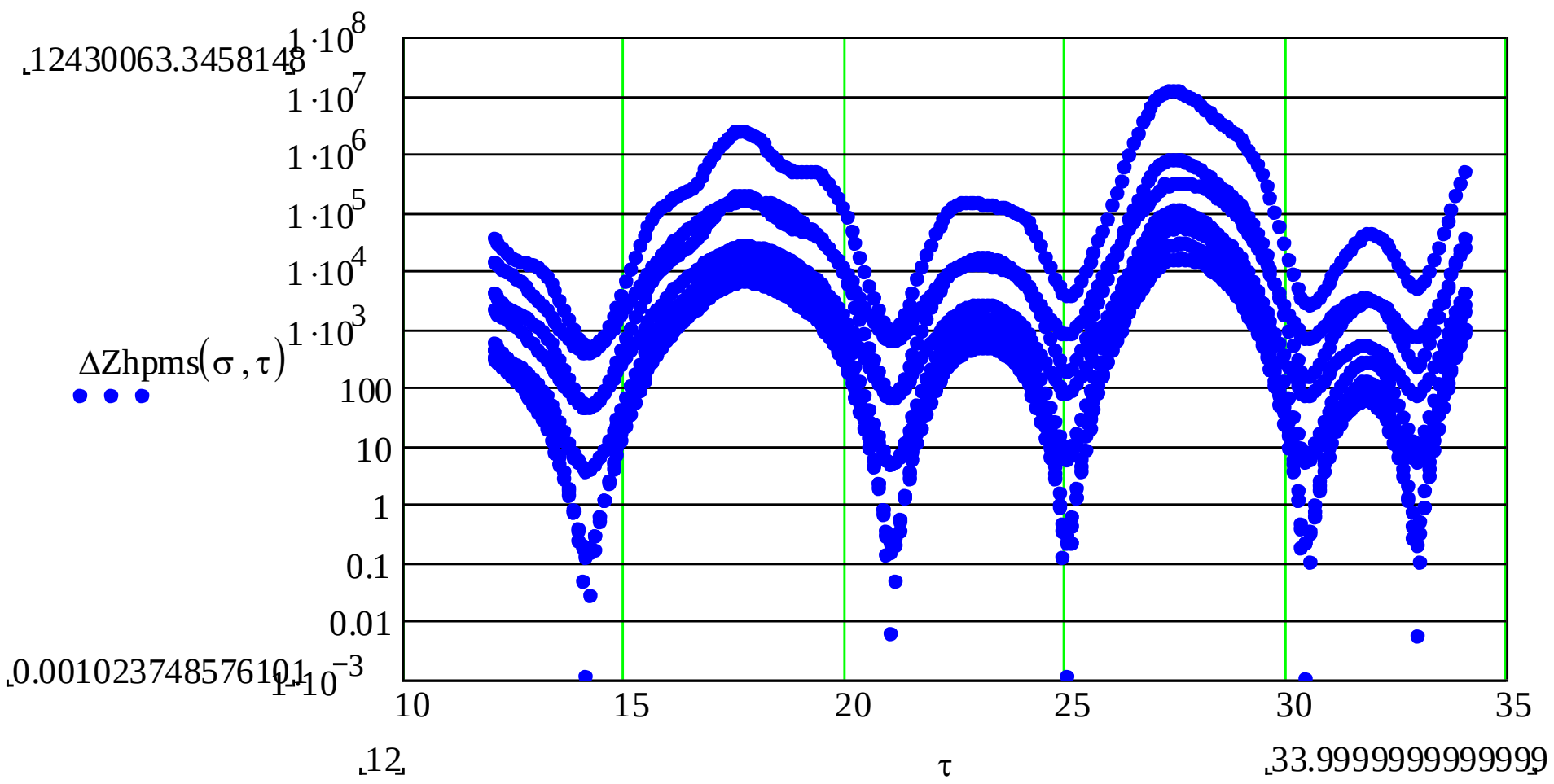

**Figure 14** Minimum-squared difference *ΔZhp,ms* versus ordinate $\tau$ varying in the interval *[10,35]*, wen varying simultaneously both $\sigma$ and $\tau$**.** Minimae values are obtained for Riemann zeroes at the critical line ordinates: $\tau$ *=14.158, 21.015, 25.000, 30.464*, and *32.948* (with a maximum relative error 0.129 % *w.r.t.* currently accepted values [28])

This plot shows that difference $\Delta Z_{hp,ms}$ to have minimae precisely at the non-trivial zeroes of RZF in the interval *[10,35]* of the critical line, as has been also found above (Section 4) when applying the AdI formalism, of course when plotting the respective variables in that case.

## 5. RIEMANN ZEROES FROM APPLYING THE LAP TO THE PARAMETRIC OSCILLATOR

In this section the LAP [13, 23-26] is applied to the parametric oscillator, this is done after warranting that it is subjected to a perturbation that is nil at the extremes of the interval time used for the evaluation of the very relevant Action integral, a *functional* as explained by Susskind and Hrabovsky [24]. In what concern to the variational function to be employed for finding the Action minimae, we shall be using the modulus of the RZF, again denoted $S(\sigma,\tau)$, over the complex plane. We begin recalling the expression of the perturbed coordinate $X_{hp}$ of this oscillator in Eq. (30) above.

$$X_{Hp}(\, t, \sigma, \tau\, , \lambda, P) \equiv X_h(t, \lambda, P) \left\{1 + S(\sigma, \tau) \sin\left[\left(\frac{t}{T}\right)\right]\right\} \qquad (39)$$

where the same variational perturbation functional of the modulus of RZF is used (see Sections 3 and 4). The kinetic energy and the potential energy of the oscillator were already given in Eqs. (12 ), (13 ) above. The Lagrangian for the perturbed oscillator is:

$$L(t; \sigma, \tau\, , \lambda, T\,) = E_k(\, t; \sigma, \tau\, , \lambda, T\,) - E_p(\, t; \sigma, \tau\, , \lambda, T\,) \qquad (40)$$

while its Action is given by the integral:

$$J(\sigma,\tau,T,\lambda) = \int_0^T L(t;\sigma,\tau,\lambda,T)dt \tag{41}$$

In Fig. 15 we have plotted this Action $J$ of the perturbed parametric oscillator as a function of the abscissa $\sigma$ in the interval (0.1, 0.9), and simultaneously varying both complex plane coordinates *($\sigma,\tau$)* in the intervals (0.1, 0.9), (12,34 ) respectively. The vertical bars of points (in blue) are due to the different values of coordinate $\tau$ being plotted as $\sigma$ varies. Note that the infimum is obtained for *$\sigma$=0.5* in the critical strip.

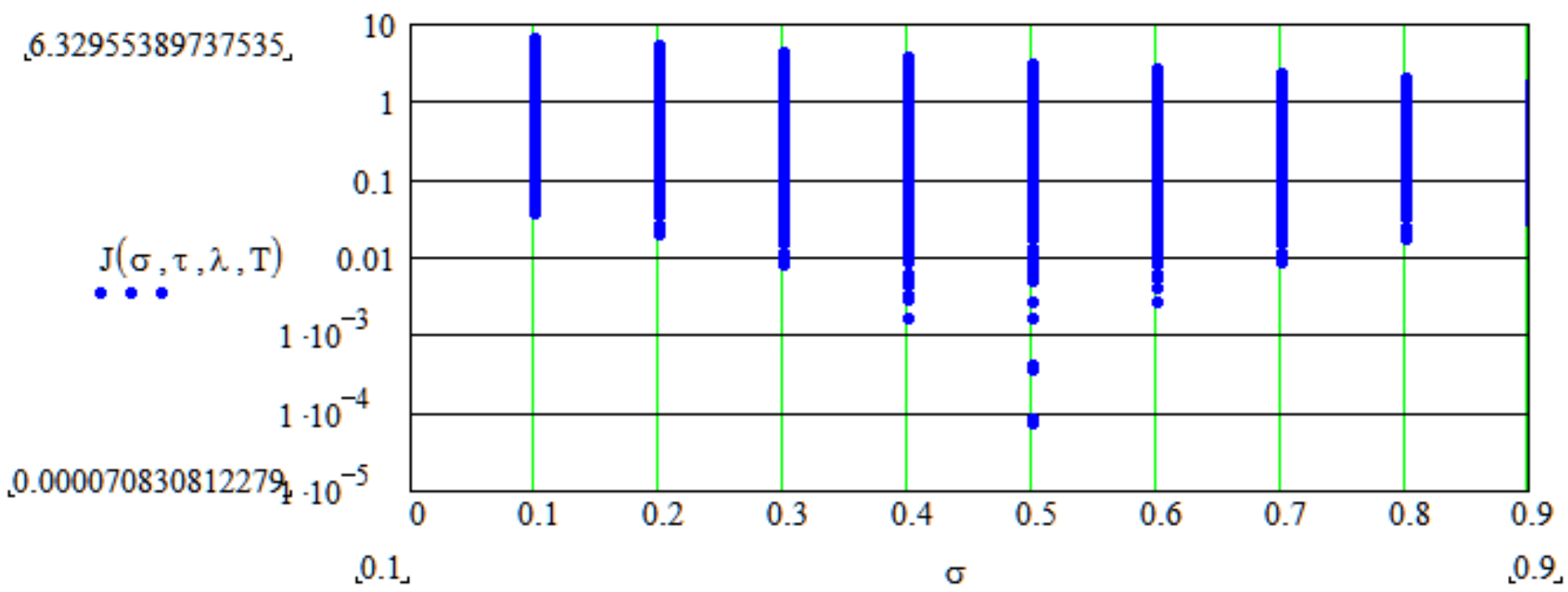

**Figure 15** Action $J$ plotted against the abscissa $\sigma$ in the critical strip (0.1,0.9) when simultaneously varying both coordinates *($\sigma,\tau$)*. Infimum of the Action points to the abscissa $\sigma_c = 0.5$ of the critical line.

Finally, in Fig. 16 we plot the action $J$ versus ordinate $\tau$ of the critical line, while varying both complex coordinates; we again get a set of minimae at the Riemann zeroes in the finite interval *[12,34]*.

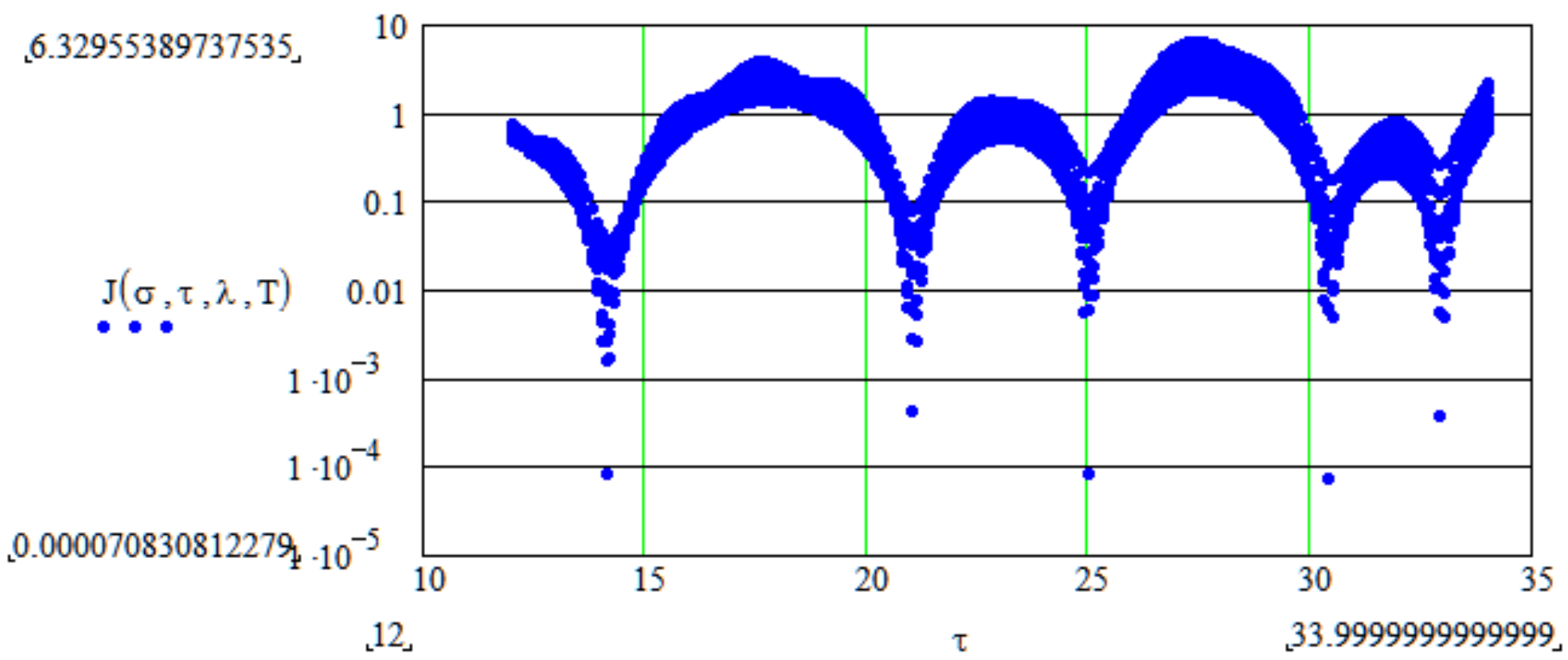

**Figure 16** Action $J$ plotted against ordinate $\tau$ in the interval *[12, 34]* along the critical line when simultaneously varying both coordinates ($\sigma$ ,$\tau$). Minimae values are obtained for Riemann zeroes at the critical line ordinates: $\tau$ *=14.114, 21.023, 25.032, 30.411*, and *32.943* (with a maximum relative error 0.046 % *w.r.t.* currently accepted values [28])

As shown in the last two graphs, the Action of the perturbed parametric oscillator *vs*. abscissa $\sigma$ in the critical strip shows a minimum at *$\sigma c$=0.5*, and when plotted *vs*. ordinate $\tau$ in a finite interval of the critical line it presents its set of minimae at the non-trivial zeroes of the RZF in that particular interval, similar to what happened above when we applied the AdI and HEF concepts to the perturbed parametric oscillator, and represented these two formalisms results versus coordinates $(\sigma,\tau)$ of the complex plane.

## 6. RIEMANN ZEROES FROM SIMULTANEOUS APPLICATION OF *LAP* TO A SET OF PERTURBED PARAMETRIC OSCILLATORS OF INTEGER PARAMETERS DEFINED USING GOLDBACH CONJECTURE

Above, we solved the motion equation of a perturbed parametric oscillator of time-variable frequency separately applying three well-known important physics-mathematical tools. In each case the perturbation of the oscillator position was defined as a functional of the modulus $S(\sigma,\tau)$ of RZF; and each time we managed to apply one of those tools, *e.g.* the Action of the oscillator, whose dependence upon the coordinates $(\sigma,\tau)$ led to the expected Riemann heights $\tau$ in finite intervals of the critical line. The variable frequency of that perturbed parametric oscillator was initially defined in terms of a positive integer parameter $\lambda$. When the fundamental condition for AdI of the perturbed oscillator was optimized (Section 3, Eqs.(16), (17)) the proper value of that parameter, namely $\lambda_c=2$ , was found in an auto-consistent way, thenceforth allowing Riemann zeroes from the perturbed oscillator to be found in either of the three cases (AdI, HEF, and Action)

Important, and truly significant, to this work shall be to develop a completely independent analytic procedure to get that proper value of integer parameter $\lambda$, or even better to replace it by an *integer-valued function* $\lambda(\sigma)$. Thus, in this section we shall find such integer-valued function. This will amount to be considering simultaneously a whole set of parametric oscillators of different $\lambda$´*s*: one oscillator for each value of coordinate $\sigma$ in the critical strip, *i.e.* a set of perturbed *test oscillators* that depend on the abscissa σ along the critical strip. This should eventually led us to find which of those oscillators generate, say Action minimae (or optimum AdI state, or optimum quality index $Q_0$ for the HEF) when the perturbation be applied to the oscillator. Actually, this should constitute a systematic method to find Riemann zeroes in finite intervals, and the critical abscissa $\sigma_c$, in summary a new way for studying Riemann´s conjecture.

To define that integer-valued function $\lambda(\sigma)$ we resort to the well-known Goldbach Conjecture for prime integer numbers, particularly written in terms of random *twin primes*; numbers that we may generate using our *prime numbers generating function* $\Psi$ defined in Eq. A2-5 of Appendix 2 [27]. These primes we write in terms of the reciprocal $(1/\sigma)$ of the abscissa $\sigma$:

$$\lambda(\sigma) \equiv 2 + \left\{2\, trunc\left[\left(\frac{1}{\sigma}\right)^2\right] - \left[\Psi\left[trunc\left[\left(\frac{1}{\sigma}\right)^2\right] - 1\right] + \Psi\left[trunc\left[\left(\frac{1}{\sigma}\right)^2\right] + 1\right]\right]\right\} \quad (42)$$

By definition the possible values of $\lambda$ should be *positive* integers (Eq. (2) of Section 2), thus Eq. (42) being written fin terms of the two *twin primes* $\left[trunc\left[\left(\frac{1}{o}\right)^2\right] - 1\right]$ , $\Psi\left[trunc\left[\left(\frac{1}{o}\right)^2\right] + 1\right]$, is a convenient and useful definition since it must be recalled that *twin primes* occurs *randomly* along the set of integers $\mathcal{N}$. In Section 3 we found a proper value $\lambda$*=2* for the perturbed oscillator parameter, therefore in Eq. (42) we are looking precisely for possible integer values of $\lambda$ *being different from 2*. Functions such as $\lambda(\sigma)$, and not particular values of parameter $\lambda$, are needed for the sake of rigour of our formalism when re-evaluating below the Lagrangian and Action functions of the perturbed oscillator.

Figure 17 we plot of the random integer function $\lambda(\sigma)$ *vs.* abscissa $\sigma$ values in the critical strip (in 0.05 discrete steps), giving us multiple possible integer values for the parameter $\lambda$. The plot shows that for $\sigma=0.5$ one gets the ordinate value $\lambda=2$.

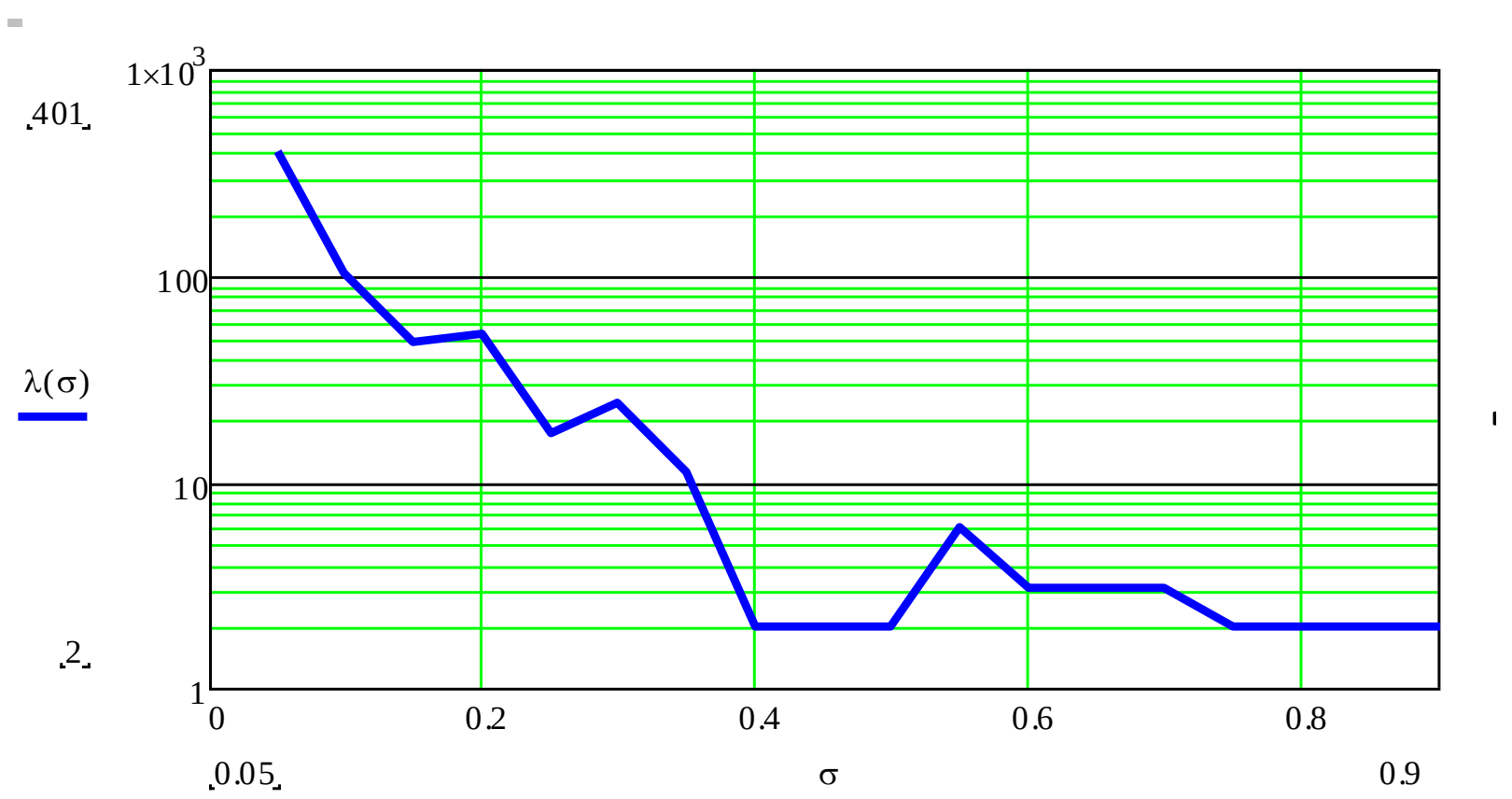

**Figure 17** Graph of the integer-valued function $\lambda(\sigma)$ versus abscissa $\sigma$ in the critical strip ($\sigma$ in 0.05 discrete steps) giving 20 possible integer values of parameter $\lambda$. Note that $\lambda(0.5)=2$.

We may now write a new Lagrangian *Lg*, and a new Action integral $J_g$ , of the perturbed parametric oscillator, by replacing the parameter $\lambda$ with the integer-valued function $\lambda(\sigma)$ in their previous expressions, Eqs.(39) and (40) above, to get respectively:

$$L_g(t;\sigma,\tau,T) = E_k(t;\sigma,\tau,\lambda(\sigma),T) - E_p(t;\sigma,\tau,\lambda(\sigma),T) \,, \tag{43}$$

and

$$J_g(\sigma,\tau,T) = \int_0^T L_g(t;\sigma,\tau,T)dt. \tag{44}$$

This Action function, evaluated using the integer-valued function $\lambda(\sigma)$, is plotted below – notably varying both coordinates $\sigma$ and $\tau$ simultaneously– versus abscissa $\sigma$ along the critical strip in Fig. 18. And once again, as in Fig. 15, the Action $A_g$ reaches its infimum value at the critical abscissa $\sigma_c= 0.5$. Note that this procedure amounts to evaluating the Action, and plotting it, using simultaneously a set of multiple *test oscillators* each of different parameter$\lambda$, yet our formalism is still capable of giving us a rather *sharp* infimum of Action $J_g$ at the critical abscissa $\sigma_{c.}=0.5$; this is indeed, a significant result.

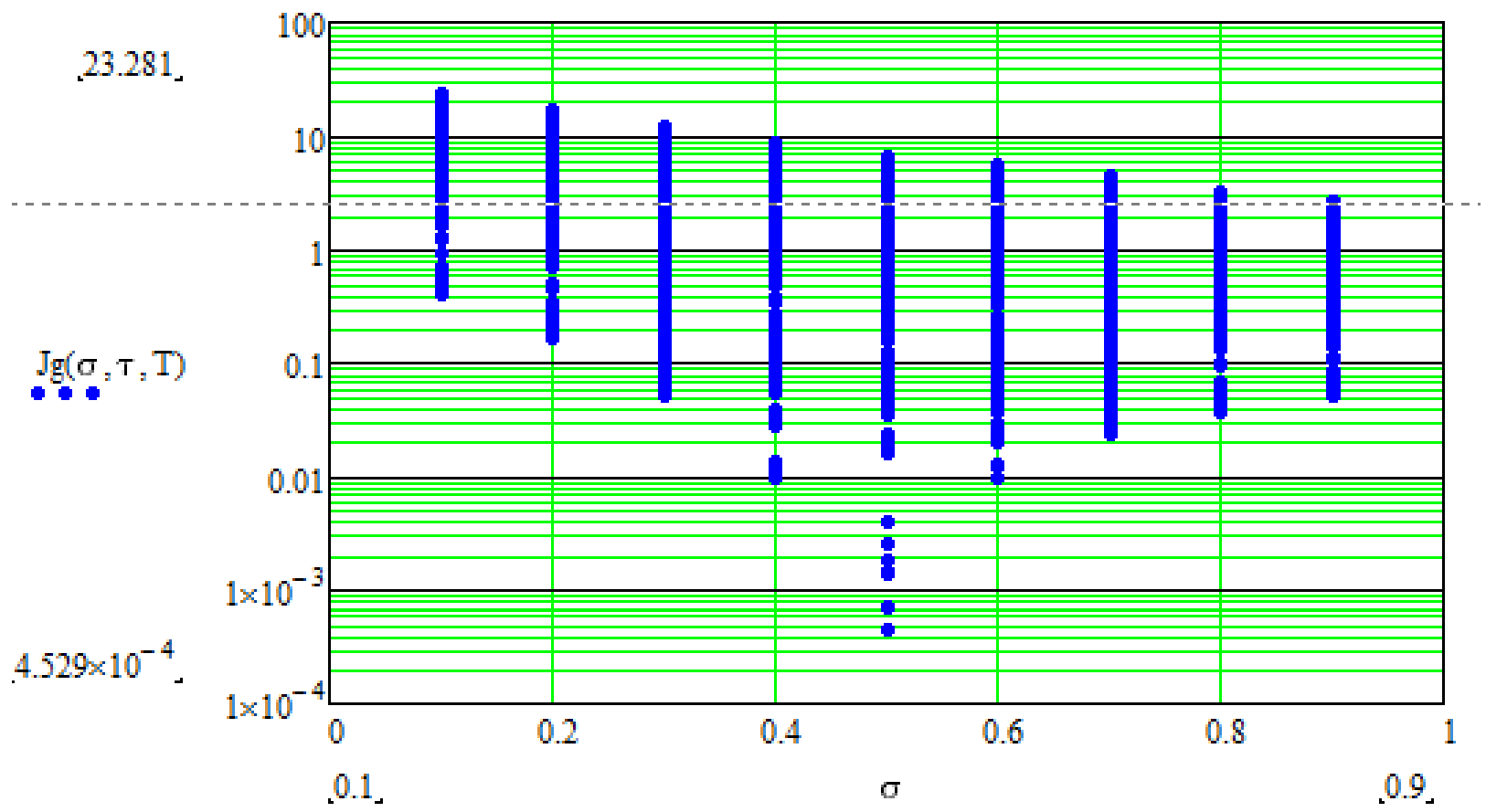

**Figure 18** Action $J_g$ of the perturbed oscillator, evaluated using the function $\lambda(\sigma)$, with both coordinates, $\sigma$ and $\tau$, varying simultaneously, plotted *vs.* abscissa $\sigma$ in the critical strip, showing the expected minimum at the abscissa $\sigma_c=0.5$ of the critical line

In Fig. 19, Action integral $J_g$ is plotted *vs.* coordinate $\tau$ values that pertain tis time to the finite interval *(90,105)* of the critical line, once again varying simultaneously both coordinates of the complex plane, *c.f.* Fig. 16. Notably, Fig. 19 shows the action minimae occurring at the Riemann zeroes in that particular interval. Again note: this time the Action has been evaluated using the integer-valued function $\lambda(\sigma)$, not any single value of the parameter λ, but simultaneously for multiple different λ values, and we still do recover the Riemann zeroes that lie in that interval *(90,105)* of the critical line .

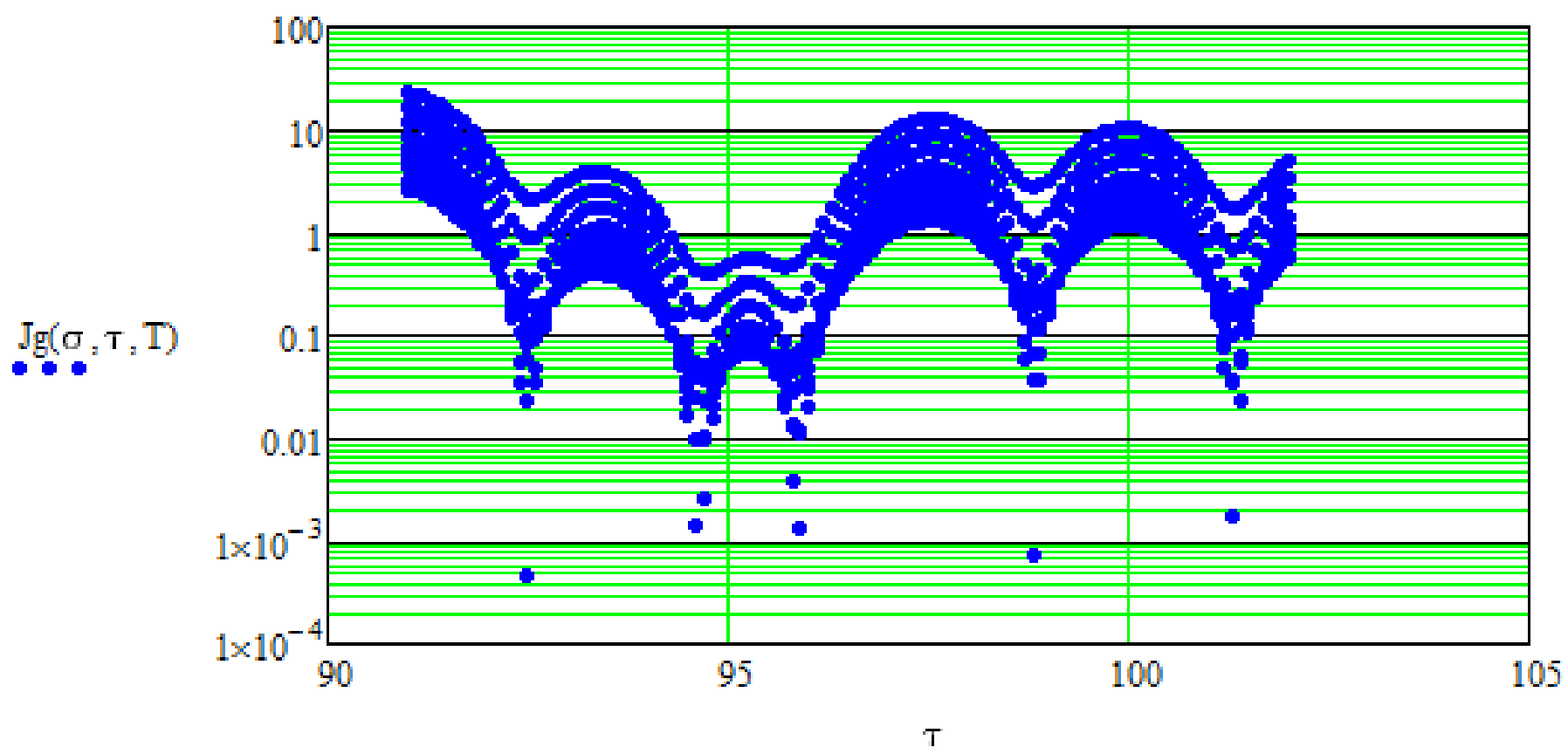

**Figure 19** Action $J_g$ of the perturbed oscillator, evaluated using function $\lambda(\sigma)$, plotted *vs* ordinate $\tau$ along the interval *(90,105)* of the critical line, with both coordinates $(\sigma,\tau)$ varying simultaneously. The plot shows minimae at the Riemann zeroes in that particular interval

We also consider convenient, just for the sake of re-assurance, to evaluate the *mean* total energy $E_g$ of the perturbed oscillator, when both the kinetic and potential energies are evaluated using the integer-valued function $\lambda(\sigma)$ defined in Eq. (42), instead of any single value of $\lambda$. The perturbed oscillator total energy is time-dependent and now given by:

$$E(t;\sigma,\tau,T) = E_k(t;\sigma,\tau,\lambda(\sigma),T)+E_p(t;\sigma,\tau,\lambda(\sigma),T)\ , \qquad (45)$$

while its *mean total* energy is:

$$E_g(\sigma,\tau,T) = 1/T\int_0^T E(t;\sigma,\tau,T)dt \qquad (46)$$

The mean energy $E_g$ evaluated using the integer-valued function $\lambda(\sigma)$ in Eq. (41), is plotted in Fig. 20 *vs*. abscissa $\sigma$ in the critical strip, while varying both coordinates $\sigma$ and $\tau$ simultaneously. Note that $E_g$ reaches its infimum value at the critical abscissa $\sigma_c = 0.5$. This being equivalent to evaluating that energy, and plotting it, using a set of *test oscillators* of many different parameters $\lambda(\sigma)$, yet remarkably the formalism is still capable of giving us a notable infimum of total energy at the critical abscissa $\sigma_c$.

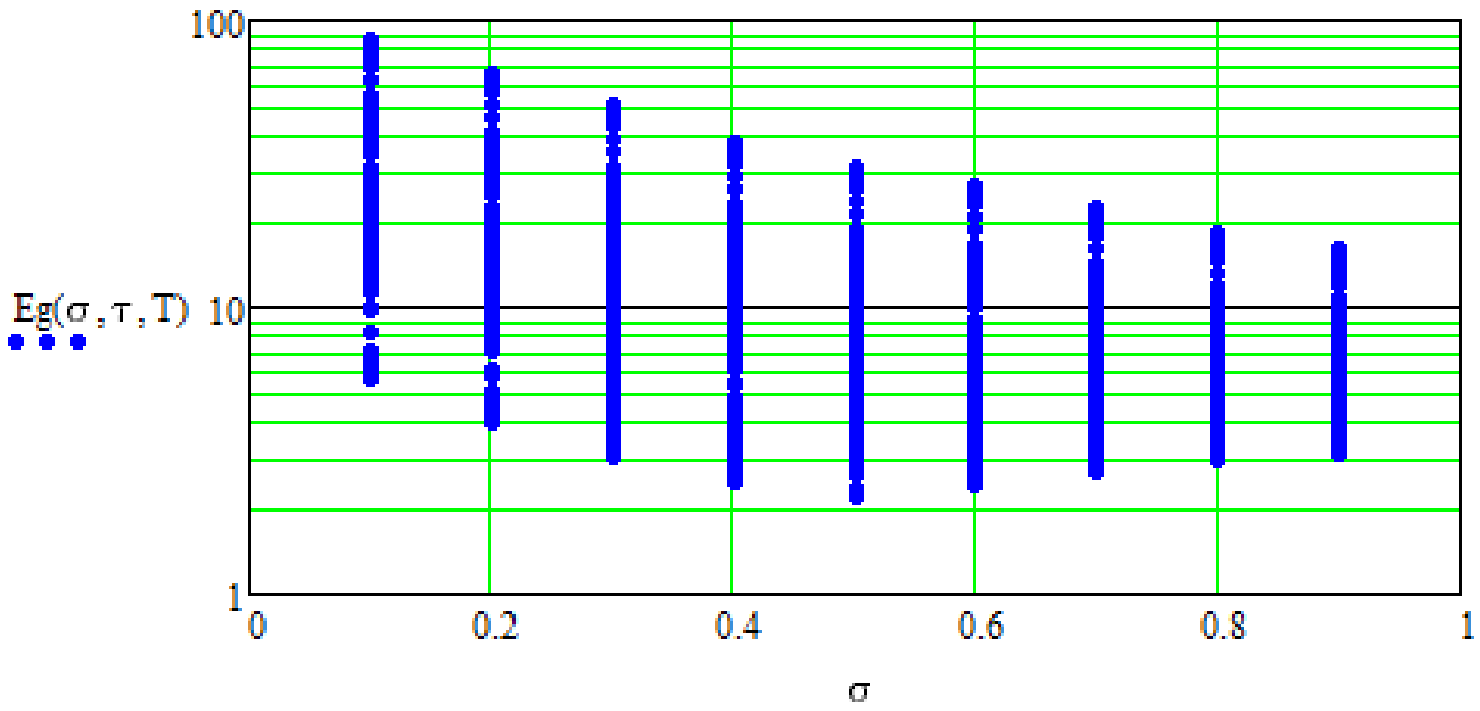

**Figure 20** Mean energy $E_g$ of the perturbed oscillator- evaluated using function $\lambda(\sigma)$ appears plotted *vs.* $\sigma$ , in the interval *(0, 1)*, showing an infimum at the abscissa critical value $\sigma_c=0.5$; plotted with both coordinates, $\sigma$ and $\tau$, varying simultaneously

In Fig. 21 the mean energy $E_g$ of the perturbed oscillator is now plotted *vs* coordinate $\tau$ that belongs to the finite interval *(90,105)* of the critical line, again varying simultaneously both coordinates of the complex plane. The plot shows the total mean energy reaching minimae at the Riemann zeroes located in that interval. Recall that the mean energy values have been evaluated using the random integer-valued function $\lambda(\sigma)$, not any single particular $\lambda$ value.

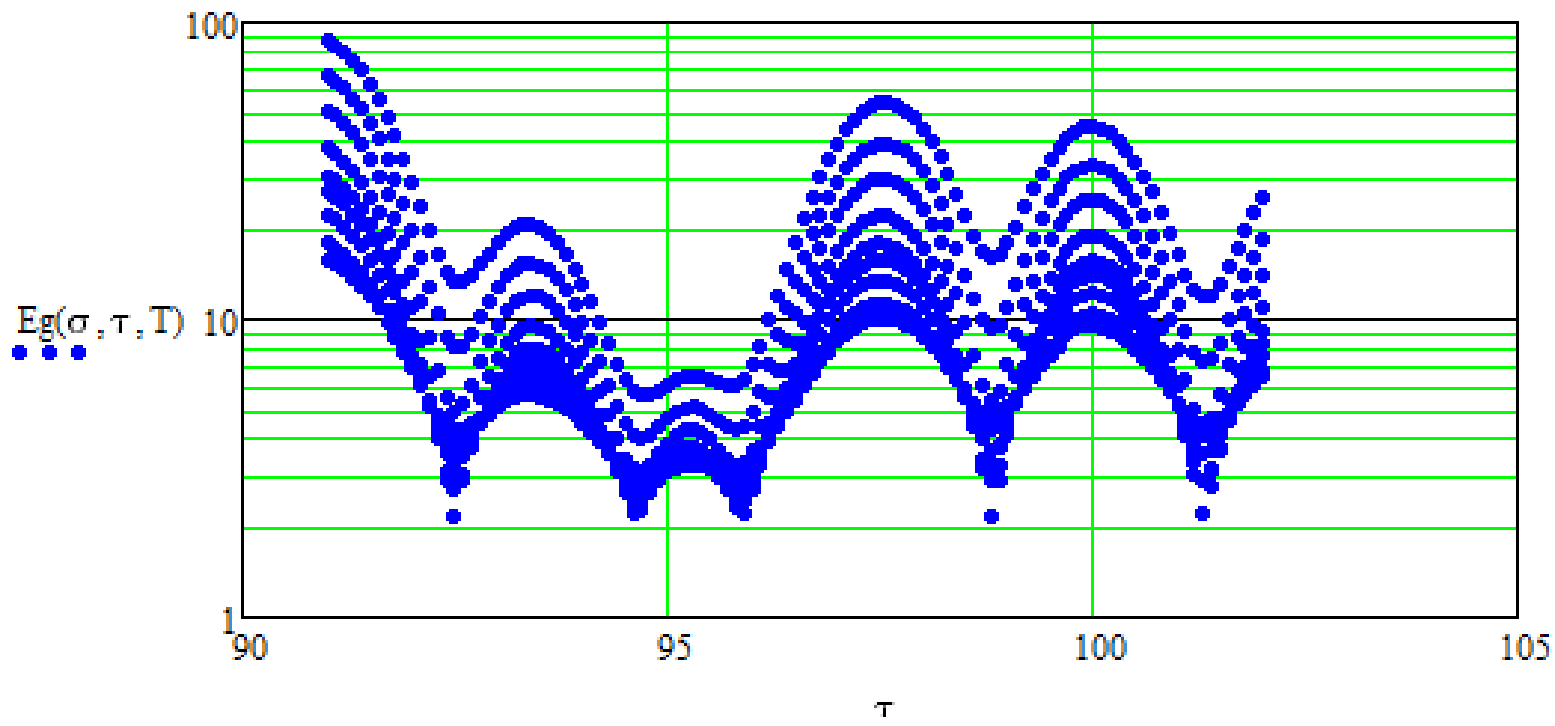

**Figure 21** Mean energy $E_g$ of the perturbed oscillator plotted *vs* ordinate $\tau \in (90,105)$ of the critical line, both coordinates $(\sigma,\tau)$ varying, and using the integer-valued function $\lambda(\sigma)$; showing minimae at the expected Riemann zeroes in that interval

Note that the calculation of the Action $(J_g)$ and the mean energy $(E_g)$ of the set of perturbed oscillators –using the random integer function $\lambda(\sigma)$– show that there exists a particular abscissa value $\sigma_c= 0.5$ of the critical strip for which both, $J_g$ and $E_g$ reach infimum values. The proper value of the parameter $\lambda=2$ (initially found in Fig. 6) is confirmed in the plot of function $\lambda(\sigma)$ in Fig. 17 for the abscissa $\sigma=\sigma_c= 0.5$. All these results are consistent with the results already obtained sections above.

## 7. CONCLUDING REMARKS

The motion equation of a one-dimensional parametric oscillator of time-variable angular frequency was solved when the oscillator motion was perturbed with a perturbation purposefully defined as a functional of the modulus $S(\sigma,\tau)$ of RZF. Three relevant tools of theoretical physics were then successfully applied to study the oscillator motion hoping to get perturbed states related to the RZF zeroes in finite intervals of the critical line. This was successfully achieved by studying the fundamental relation, or concept, at the core of each of the three theoretical tools, and the conditions for that relation or concept to reach stationary or minimum values, which as expected lead precisely to Riemann zeroes.

We began applying the fundamental AdI theoretical tool (Section 2), as presented by Landau and Lifschits [13], that demands the quotient between the total energy of an oscillator and its angular frequency to be a constant of motion when the adiabatic regime is achieved. It led us to define and evaluate a *quality index* $Q_0$ capable of qualifying any state of adiabatic invariance in the motion of our oscillator. In fact this index reached the expected minimae precisely at the Riemann zeroes in arbitrarily chosen intervals of the ordinate of the critical line, zeroes found with good accuracy. When the index $Q_0$ was evaluated as a function of the abscissa $\sigma$, and for a known ordinate $\tau_0$ of a particular known Riemann zero, the index reached its minimum value precisely at the abscissa $\sigma_0=0.5$ of the critical line This is indeed a novel application of the AdI formalism. Better yet, when an analogous evaluation of $Q_0$ was done but allowing both coordinates $(\sigma,\tau)$ to vary simultaneously, once again and remarkably the manifold of plotted functions signalled once again the expected minimum at $\sigma_0=0.5$ of the critical line. Evaluating index $Q_0$ when both coordinates $(\sigma,\tau)$ are varied resulted in that index taking minimum values at the same Riemann zeroes, or heights, on the critical line of the RZF that have been found using other methods [8, 9].

In Section 3 we introduced a novel mathematical-physics formalism (HEF) of theoretical physics based on the well-known Hill´s differential equation, and particularly on a second order nonlinear differential equation (Eq. 25) discovered by Magnus and Winkler (M-W) [21]; it is a novel and useful method applicable to study any perturbed physical system that admits a Hill´s equation representation. This nonlinear M-W equation is obtained from two known *LI* solutions of any given Hill´s equation. Our HEF formalism was then applied to study our parametric oscillator (defined in Section 2) in both cases: when unperturbed and when perturbed with a perturbation functional of the RZF. In the latter case we derived a relative difference function $\Delta Z_{hp,rel}$ (Eq. 35) which minimum-squared $\Delta Z_{hp,ms}$ we defined (Eq. 37) and then numerically calculated. Used as a *merit figure* the quantity$\Delta Z_{hp,ms}$ allowed us to get zeroes of the RZF from the perturbed oscillator case. In effect, when $\Delta Z_{hp,ms}$ is plotted *vs*. coordinate $\tau$ it does show minimae (order of $10^{-3}$) at the known zeroes of RZF in arbitrary intervals of the critical line. Moreover, when plotted against abscissa $\sigma$ in the critical strip, the minimum of $\Delta Z_{hp,ms}$ occurred at the abscissa $\sigma_0$= 0.5 of the critical line. Next, we applied the LAP to find nontrivial zeroes from our RZF perturbed oscillator (Section 5), and we followed the well-known procedure: just find that minimae of an Action *J* defined using a Lagrangian written in terms of a perturbation functional of the modulus $S(\sigma,\tau)$ of RZF. The Action plots versus coordinates $\sigma$ and $\tau$ once again gave us minimae either at abscissa $\sigma_0$= 0.5 of the critical line, or at the expected zeroes of RZF along arbitrary finite intervals of the critical line, such minimae being as least as $10^{-4}$.

The parametric oscillator –to which the LAP, HEF and AdI theoretical tools were applied in Sections 3, 4 and 5 – has a variable frequency that depends upon a positive integer parameter $\lambda$, and whose value $(\lambda=2)$ was later found (Fig. 6, Section 3). However, in Section 5 we successfully solved the problem of applying the LAP to the perturbed oscillator *without requiring the determination of any particular value of parameter*$\lambda$. It was done by replacing that parameter $\lambda$ in the Lagrangian perturbation by a random integer-valued function $\lambda(\sigma)$ that we defined using Goldbach Conjecture written in term of random twin primes, just to warrant an unbiased function $\lambda(\sigma)$. This function we evaluated, and plotted in Fig. 17, for abscissa $\sigma$ values in the interval (0.1, 0.9) and increasing in 0.05 steps, giving multiple different integer values of parameter $\lambda$. The Action of the perturbed oscillator was calculated, and plotted anew replacing the parameter $\lambda$ by the function $\lambda(\sigma)$ (in Section 5). Even when varying both coordinates $(\sigma,\tau)$ and calculating with that integer-valued function, the Action ($J_g$) once again showed the minimae at the Riemann zeroes along finite intervals of the critical line (Fig.

19) and its infimum at the critical abscissa $\sigma_C=0.5$ Plotting the mean total energy of the oscillator varying both coordinates $(\sigma,\tau)$, and replacing $\lambda$ by the integer-valued function $\lambda(\sigma)$ showed analogous results, proving that the minimae and infimum obtained for the Action ($J_g$) was not a fortuitous result.

The three physics-mathematical formalisms here presented are rigorous analytical verifications that Riemann Conjecture must hold at least for the nontrivial zeroes found on those finite intervals of the critical line used in our plots. Results in this work were numerically obtained using an approximate expression for the RZF (Eq. 1) for $N=1000$, and using a modest 16 bytes PC computer. The ordinates of the Riemann zeroes that we have obtained, in the interval *[12, 34]* of the critical line, are listed in the legends of Figs. 8, 14 and 16, with maximum relative errors: 0.039 %, 0.129 %, 0.046 %, respectively. Certainly, much better and accurate results for the Riemann zeroes would be obtained with our three formalisms using a more powerful computer and larger $N$ values in the expression for RZF given in Eq (1).

**ACKNOWLEDGEMENTS**

We acknowledge the support of Decanato de Investigaciones y Desarrollo**,** Universidad Simón Bolívar, Caracas, Venezuela.

**APPENDIX 1**

Here we present the kinematics of the parametric oscillator introduced in Section 2, and validate it presenting the method of solution of its Hill´s equation that lead to the pair of *LI* solutions in that section. Initially one may write the first linearly independent solution $X$ as:

$$X(t) = R(t)\cos[(\theta(t)] , \qquad \text{(A1-1)}$$

where $\lambda,T$ are the parameters of our model in that section. After derivation of this equation we get:

$$\frac{d}{dt}X(t) = \frac{dR(t)}{dt} cos\,[(\theta(t)] - R(t)\, sin[(\theta(t)]\frac{d}{dt}[(\theta(t)] = \frac{X(t)}{R(t)}\frac{d}{dt}R(t) - R(t)\, sin[(\theta(t)]\frac{d}{dt} sin[(\theta(t)],$$

where it is convenient to introduce the functions $\omega$ and $\eta$ :

$$\omega(t) = \frac{d\theta(t)}{dt} \qquad \text{(A1-2)}$$

$$\eta(t) = \frac{1}{R(t)}\frac{d}{dt}R(t), \qquad \text{(A1-3)}$$

Note $\omega$ is the initial key input for later defining the variable frequency $\Omega$ of the parametric oscillator, while function $\eta$ is just an auxiliary function with frequency dimensions. Let us now calculate the second derivative of coordinate $X$ ,

$$\frac{d^2}{dt^2}X(t) = \left[\frac{d}{dt}\eta(t) + \eta^2(t) - \omega^2(t)\right]X(t) - \left[2\omega(t)\eta(t) + \frac{d}{dt}\omega(t)\right]R(t)\sin[(\theta(t)]$$

Where we define the variable frequency $\Omega$ as:

$$\Omega^2(t) = \omega^2(t) - \left[\frac{d}{dt}\eta(t) + \eta^2(t)\right] = \omega^2(t)\left\{1 - \frac{1}{\omega^2(t)}\left[\frac{d}{dt}\eta(t) + \eta^2(t)\right]\right\}$$

where we have rewritten the function $\eta$ as:

$$\eta(t) = \frac{-\frac{d}{dt}\omega(t)}{2\omega(t)}, \qquad \text{(A1-4)}$$

since it is well-known that the wronskian of Hill´s equation is a constant. Thus we may rewrite Hill´s equation for the parametric oscillator abscissa $X$ as:

$$\frac{d^2}{dt^2}X(t;\, \lambda,, T) + \Omega^2(t)X(t;\, \lambda,, T) = 0,$$

Once the function $\omega$ be defined, as done in Eq. (3 ) of Section 2, and using (A4) the analytic solution of Hill´s equation for the position coordinate *X(t)* of the parametric oscillator is:

$$X(t) = R_i exp\left[\int_0^t \eta(t')dt'\right]\left\{\left[\cos\left(\theta_i + \int_0^t \omega(t')dt'\right)\right]\right\}, \qquad \text{(A1-5)}$$

where *Ri* is the initial coordinate.

Analogously, the second linearly independent solution $Y$ is given by:

$$Y(t) = R_i exp\left[\int_0^t \eta(t')dt'\right]\{[\sin(\theta_i + \int_0^{\prime} \omega(t')dt')]\} \qquad \text{(A1-6)}$$

**APPENDIX 2**

To define our Prime Numbers Generating function [26, 27] we first construct the following three auxiliary functions $\Omega_i$ of $u$, and integer's $m$ and $n$, in terms of the well-known function $\eta(u,u_0) = sign(u-u_0)$:

$$\Omega_0(u) = [\eta(|u|, 0)]^2[\eta(|u|, 1)]^2[1 - \eta(\Delta(u), 0)] \qquad \text{(A2-1)}$$

$$\Omega_1(u, m) = [\eta(|u|, 2m)]^2 \,, \qquad \text{(A2-2)}$$

$$\Omega_2(u, m, n) = \{\eta[|u|, (2m-1)(2n-1)\ ]\}^2, \qquad \text{(A2-3)}$$

where $m > 2, n > 2$. With these three functions we define a *Prime Numbers Discriminating* function, denoted $\Lambda$, as the double-product:

$$\Lambda(u) \equiv \Omega_0(u) \prod_{m=2}^{h_1(u)} \prod_{n=2}^{h_2(u,m)} [\Omega_1(u,m)\Omega_2(u,m,n)] \, . \qquad \text{(A2-4)}$$

Prime Numbers Generator function $\Psi$ may now be defined, in terms of the Prime Numbers Discriminating function $\Lambda$, as:

$$\Psi(u) \equiv u\Lambda(u) \; , \qquad \text{(A2-5)}$$

that appears plotted in Fig. A2-1 for $u \in [0,100]$, prime numbers along the inclined line

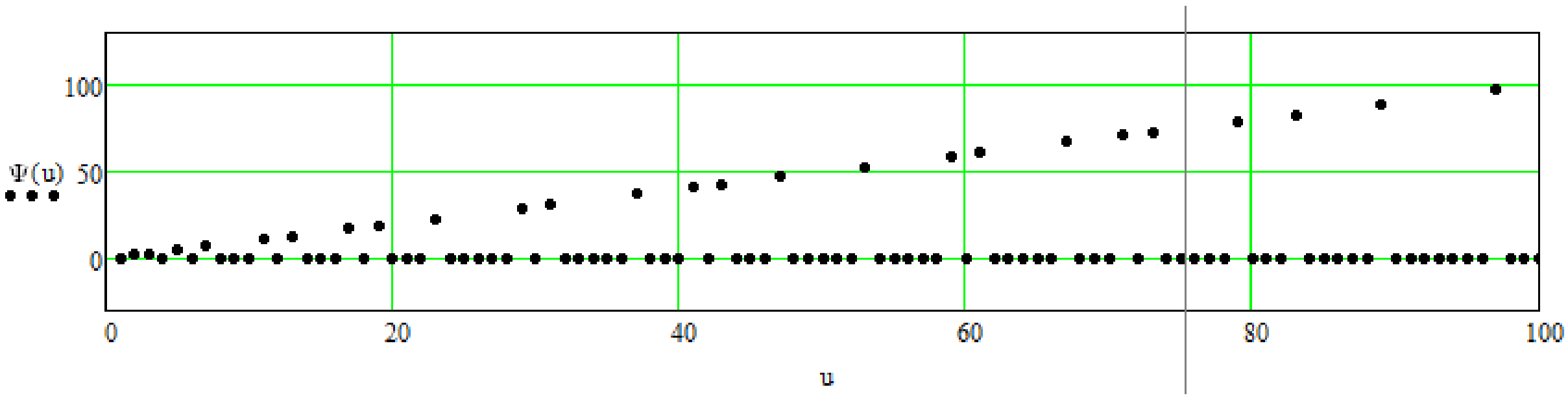

**Figure A2-1** Graph of the Prime Generating Function $\Psi$ for $u \in [0,100]$: prime numbers plotted as dots along the inclined line.

## REFERENCES

[1] van der Pol B 1947 An electro-mechanical investigation of the Riemann zeta function in the critical strip , *Bull. Am. Math. Soc.* **53** 976–81

[2] Berry M V and Keating J P**,** The Riemann Zeros and Eigenvalue Asymptotics, *SIAM REV*, **41**, No. 2, pp. 236–266

[3] Sierra G, 2007 H = xp with interaction and the Riemann zeros, arXiv: math ph/0702034v2

[4] Mack R, Dahl J P, Moya-Cessa H, Strunz W T, Walser R and Schleich W P 2010 Riemann zeta function from wave-packet dynamics, *Phys. Re*v. A82, 032119.

[5] Planat M, Sol´e P and Omar S, Riemann hypothesis and quantum mechanics, 2011 arXiv 1012-4665 **3** [math ph]

[6] Berry M V 2012 Riemann zeros in radiation patterns, *J. Phys. A: Math. Theor.* **45**, 302001

[7] Feiler C and Schleich W P 2013 Entanglement and analytical continuation:an intimate relation told by the Riemann zeta function, *New J. Phys.* **15** 063009

[8] Berry M V 2015 Riemann zeros in radiation patterns II: Fourier transform of Zeta, *J. Phys. A: Math. Theor.* **48**, 385203

[9] Ran H, Ai M-Z, Cui J-M, Huang Y-F, Han Y-J, Chuan-Feng L, Tao T, Creffield C E, Sierra G, and Guo G-C 2020 Identifying the Riemann zeros by periodically driving a single qubit, *Phys. Rev. A* **101**, 043402

[10] Mussardo G and Leclair A 2021 Randomness of Möbius coefficients and Brownian motion: growth of the Mertens function and the Riemann hypothesis, *J. Stat.Mech: Theory and Experiments*

[11] Chen G, Guo G, Yang K, D. Yang D 2021 Multi-step prediction of zero series and gap series of Riemann zeta function, *Results in Physics* 27 104449

[12] Ehrenfest P 1916 On adiabatic changes of a system in connection with the quantum theory, *Koninklijke Akademie van Wetenschappen te Amsterdam* Section on Sciences Proceedings **19** 576-597

[13] Landau L D, Lifshitz E M 1976 *Mechanics*, *Course of Theoretical Physics*, Vol. 1, (New York, Pergamon)

[14] Kulsrud R M 1957Adiabatic Invariant of the Harmonic Oscillator, *Phys. Rev.* 106**, 205**

[15] Anosov D V, Favorskii A P 1988 Adiabatic invariant, *Encyclopaedia of Mathematics*, Volume 1 (Dordrecht, Reidel)

[16] Cap F F 1978 Adiabatic invariants and motion in special fields, Handbook on Plasma Instabilities, Vol 1, Ch. 14 (New York, Academic Pr)

[17] Hill G W 1886 *On the part of the motion of the lunar perigee which is the function of mean motions of the sun and* moon, *Acta Math*, **8**, 1-36

[18] Arnold V I 1989 *Mathematical Methods of Classical Mechanics*, Springer

[19] Jordan D W and Smith P 1977 *Nonlinear Ordinary Differential Equations: An introduction for Scientists and Engineers*, 4th. ed. (Oxford, Oxford Univ Pr)

**[20]** Humi M and Miller W 1988, *Second Course in Ordinary Differential Equations for Scientists and Engineers* (New York , Springer-Verlaj)

[21] Magnus W and Winkler S 1966, *Hill´s Equations* (New York, Interscience,)

**[22]** Cap F 1978 Parametric effects, *Handbook on Plasma Instabilities*, Vol 2, Ch.17 (New York, Academic Pr.)

[**23]** Feynman R P, Leighton R S and Sands M 1964, *Feynman Lectures on Physics* **Vol. 2** Lecture 19 (New York, Addison Wesley)

**[24**] Susskind P and Hrabovsky G 2012, *The theoretical minimum,* (New York, Penguin Books)

**[25]** Taylor E F 2003 A Call to Action, *Am. J. Phys*. 71 (5) 423-425.

[26] Stella E and Ladera C L 2021, New generating and counting Functions of prime numbers applied to approximate Chebyschev 2nd class function and the least action principle applied to find non-trivial roots of the Zeta function and to Riemann Hypothesis, arXiv:2106.10228v4 [math.GM]

[27] Stella E, Ladera C L and Donoso G 2022 New Generating and Counting Functions of Prime Numbers applied to Bertrand Theorem, Euler´s Product formula, Prime Numbers Products and Chebyschev $2^{nd}$. class function hal-03527940

[28] http://www.dtc.umn.edu/~odlyzko/zeta_tables/index.html

[29] Moree P I, Petrykiewicz P I, and Sedunova A 2018 A computational history of prime numbers and Riemann zeros, arXiv:1810.05244v1 [math.NT]